\documentclass[aps,prl,preprint,nopacs,superscriptaddress]{revtex4}
\usepackage{graphicx}
\usepackage{verbatim}
\usepackage{mathrsfs}
\usepackage{amsmath,amsfonts,amssymb}
\usepackage{graphicx}

\def\3{2.8in}    
\def\2{2.5in}
\def\4{3.0in}

\def \beq {\begin{equation}}
\def \eeq {\end{equation}}
\begin{document}

\title{A topological crystalline insulator phase via topological phase transition and crystalline mirror symmetry}
\author{Su-Yang Xu}\affiliation {Joseph Henry Laboratory, Department of Physics, Princeton University, Princeton, New Jersey 08544, USA}
\author{Chang Liu}\affiliation {Joseph Henry Laboratory, Department of Physics, Princeton University, Princeton, New Jersey 08544, USA}
\author{N. Alidoust}\affiliation {Joseph Henry Laboratory, Department of Physics, Princeton University, Princeton, New Jersey 08544, USA}
\author{M. Neupane}\affiliation {Joseph Henry Laboratory, Department of Physics, Princeton University, Princeton, New Jersey 08544, USA}
\author{D. Qian}\affiliation {Joseph Henry Laboratory, Department of Physics, Princeton University, Princeton, New Jersey 08544, USA}\affiliation {Key Laboratory of Artificial Structures and Quantum Control (Ministry of Education), Department of Physics, Shanghai Jiao Tong University, Shanghai 200240, China.}
\author{I. Belopolski}\affiliation {Joseph Henry Laboratory, Department of Physics, Princeton University, Princeton, New Jersey 08544, USA}

\author{J. D. Denlinger}\affiliation {Advanced Light Source, Lawrence Berkeley National Laboratory, Berkeley, California 94305, USA}
\author{Y. J. Wang}\affiliation {Department of Physics, Northeastern University, Boston, Massachusetts 02115, USA}
\author{H. Lin}\affiliation {Department of Physics, Northeastern University, Boston, Massachusetts 02115, USA}
\author{L. A. Wray}\affiliation {Joseph Henry Laboratory, Department of Physics, Princeton University, Princeton, New Jersey 08544, USA}\affiliation {Advanced Light Source, Lawrence Berkeley National Laboratory, Berkeley, California 94305, USA}

\author{G. Landolt}\affiliation {Swiss Light Source, Paul Scherrer Institute, CH-5232, Villigen, Switzerland}\affiliation {Physik-Institute, Universitat Zurich-Irchel, CH-8057 Zurich, Switzerland}
\author{B. Slomski}\affiliation {Swiss Light Source, Paul Scherrer Institute, CH-5232, Villigen, Switzerland}\affiliation {Physik-Institute, Universitat Zurich-Irchel, CH-8057 Zurich, Switzerland}
\author{J. H. Dil}\affiliation {Swiss Light Source, Paul Scherrer Institute, CH-5232, Villigen, Switzerland}\affiliation {Physik-Institute, Universitat Zurich-Irchel, CH-8057 Zurich, Switzerland}
\author{A. Marcinkova} \affiliation{Department of Physics and Astronomy, Rice University, Houston, Texas 77005, USA}

\author{E. Morosan} \affiliation{Department of Physics and Astronomy, Rice University, Houston, Texas 77005, USA}
\author{Q. Gibson}\affiliation {Department of Chemistry, Princeton University, Princeton, New Jersey 08544, USA}

\author{R. Sankar} \affiliation{Center for Condensed Matter Sciences, National Taiwan University, Taipei 10617, Taiwan}
\author{F. C. Chou} \affiliation{Center for Condensed Matter Sciences, National Taiwan University, Taipei 10617, Taiwan}

\author{R. J. Cava}\affiliation {Department of Chemistry, Princeton University, Princeton, New Jersey 08544, USA}
\author{A. Bansil}\affiliation {Department of Physics, Northeastern University, Boston, Massachusetts 02115, USA}
\author{M. Z. Hasan}\affiliation {Joseph Henry Laboratory, Department of Physics, Princeton University, Princeton, New Jersey 08544, USA}\affiliation{Princeton Center for Complex Materials, Princeton Institute for Science and Technology of Materials, Princeton University, Princeton, New Jersey 08544, USA}


\begin{abstract}
\textbf{A topological insulator protected by time-reversal symmetry is realized via spin-orbit interaction driven band inversion. The topological phase in the Bi$_{1-x}$Sb$_x$ system is due to an odd number of band inversions. A related spin-orbit system, the Pb$_{1-x}$Sn$_x$Te, has long been known to contain an even number of inversions based on band theory. Here we experimentally investigate the possibility of a mirror symmetry protected topological crystalline insulator phase in the Pb$_{1-x}$Sn$_x$Te class of materials which has been theoretically predicted to exist in its end compound SnTe. Our experimental results show that at a finite-Pb composition above the topological inversion phase transition, the surface exhibits even number of spin-polarized Dirac cone states revealing mirror-protected topological order distinct from that observed in Bi$_{1-x}$Sb$_x$. Our observation of the spin-polarized Dirac surface states in the inverted Pb$_{1-x}$Sn$_x$Te and their absence in the non-inverted compounds related via a topological phase transition provide the experimental groundwork for opening the research on novel topological order in quantum devices.}
\end{abstract}
\date{\today}
\maketitle


Topological insulators are band insulators featuring uniquely spin-polarized surface states, which is believed to be a consequence of odd number of inversions of bulk bands driven by spin-orbit interaction alone or in combination with crystal lattice modulations \cite{Moore, RMP, Zhang_RMP, Hasan QPT, Matthew Nature physics BiSe}. The odd number of band inversion transitions in the bulk demarcate a Z$_2$ topological insulator phase with spin-polarized surface states from that of a conventional band insulator phase as previously demonstrated in topological insulator Bi$_{1-x}$Sb$_x$ and the Bi$_2$Se$_3$ series \cite{Moore, RMP, Zhang_RMP, Hasan QPT, Matthew Nature physics BiSe}. Within band theoretical calculations, lead tin telluride Pb$_{1-x}$Sn$_{x}$Te, which is a narrow band-gap semiconductor widely used for infrared optoelectronic and thermoelectric devices \cite{PbTe IR, PbTe Thermal}, has long been known to contain an even number of band inversions as electronic structure is tuned via the Sn/Pb ratio \cite{PST band-gap, PST Inversion1, PST Inversion2, PST Inversion3, Volkov, Fradkin}. Theoretical calculations have also predicted the occurrence of surface or interface states within the inverted band-gap upon the band inversion transition \cite{PST Inversion3, Volkov, Fradkin} but no experimental evidence of the surface states has been reported so far. Recent theoretical works \cite{Liang PRL TCI, Liang NC SnTe} have further stimulated the experimental search for surface states in Pb$_{1-x}$Sn$_{x}$Te. In a recent calculation \cite{Liang NC SnTe}, Hsieh \textit{et al}. predicted that the topological surface states in SnTe \cite{Liang NC SnTe} are protected by crystalline space group symmetries, in contrast to the time-reversal symmetry protection in the well known Z$_2$ topological insulators.


The crystal group symmetry of Pb$_{1-x}$Sn$_{x}$Te, critical for the realization of the topological crystalline insulator (TCI) phase \cite{Liang PRL TCI}, is based on the sodium chloride structure (space group Fm$\bar{3}$m (225)). In this structure, each of the two atom types (Pb/Sn, or Te) forms a separate face-centered cubic lattice, with the two lattices interpenetrating so as to form a three-dimensional checkerboard pattern. The first Brillouin zone (BZ) of the crystal structure is a truncated octahedron with six square faces and eight hexagonal faces. The band-gap of Pb$_{1-x}$Sn$_{x}$Te is found to be a direct gap located at the $L$ points in the BZ \cite{PST band-gap}., which are the centers of the eight hexagonal faces of the BZ. Due to the inversion symmetry of the crystal, each $L$ point and its diametrically opposite partner on the BZ are completely equivalent. Thus there are four distinct $L$ point momenta. It is well established that the band inversion transitions in the Pb$_{1-x}$Sn$_{x}$Te take place at these four $L$ points of the BZ \cite{PST Inversion1, PST Inversion2, PST Inversion3, Volkov, Fradkin}. As a result, even number of (four) inversions exclude the system from being a Kane-Mele Z$_2$ topological insulator \cite{RMP}. However, it is interesting to note that the momentum-space locations of the band inversions coincident with the momentum-space mirror plane within the BZ. This fact provides a clue that band inversions in Pb$_{1-x}$Sn$_{x}$Te may lead to a distinct topologically nontrivial phase that is irrelevant to the time-reversal symmetry but may be the consequence of the spatial mirror symmetries of the crystal, which has been theoretically shown in the inverted end compound SnTe \cite{Liang NC SnTe}. However, it is experimentally known that SnTe crystals actually subject to a rhombohedral distortion \cite{PST Rhombohedral Distortion}, which, strictly speaking, breaks the crystal mirror symmetries and hence excludes a gapless TCI phase \cite{Liang NC SnTe}. More importantly, SnTe crystals are typically heavily p-type \cite{SnTe p-type} due to the fact that Sn vacancies are thermodynamically stable \cite{SnTe p-type}, which makes the chemical potential of SnTe to lie below the bulk valence band maximum, consequently not cutting across the surface states. The surface states, which may exist within the band-gap, are thus unoccupied and cannot be experimentally studied by photoemission experiments \cite{SnTe ARPES}. On the other hand, the Pb-rich samples are reported to possess the ideal (averaged) sodium chloride structure without global rhombohedral distortion \cite{PST Rhombohedral Distortion} which thus preserve the proper crystal symmetry required for the predicted TCI phase \cite{Liang NC SnTe}. Moreover, it is possible to grow Pb-rich crystals, in which the chemical potential can be brought up above the bulk valence band maximum (in-gap or even n-type) \cite{PST n-type} so as to lie near the surface state Dirac point. 


Therefore, in order to experimentally investigate the possibility of a topological phase transition, and thus rigorously proving the existence of a TCI phase in the Pb$_{1-x}$Sn$_{x}$Te system, we hereby utilize angle-resolved photoemission spectroscopy (ARPES) and spin-resolved ARPES to study the low energy electronic structure below and above the band inversion topological transition in the Pb-rich compositional range of the system, through which the observation of spin-polarized surface states in the inverted regime and their absence in the non-inverted regime is demonstrated, correlating the observed spin-polarized Dirac surface states with the expected band inversion topological phase transition. We further map out the critically important spin structure of the surface states protected by mirror symmetries in the inverted topological composition. These results show that the topological order observed in Pb$_{1-x}$Sn$_{x}$Te is distinct from that previously discovered in Bi$_{1-x}$Sb$_x$ or Bi$_2$Se$_3$ \cite{Moore, RMP, Zhang_RMP, Hasan QPT, Matthew Nature physics BiSe}. Our spin-resolved study of Pb$_{1-x}$Sn$_{x}$Te paves the way for further investigation of many novel properties of topological crystalline order \cite{Liang PRL TCI, Yannopapas, Wang, Vildanov} in real materials and devices.

\bigskip
\bigskip
\textbf{Results}
\newline
\textbf{Comparison of non-inverted and inverted compositions}

\begin{figure*}
\centering
\includegraphics[width=18cm]{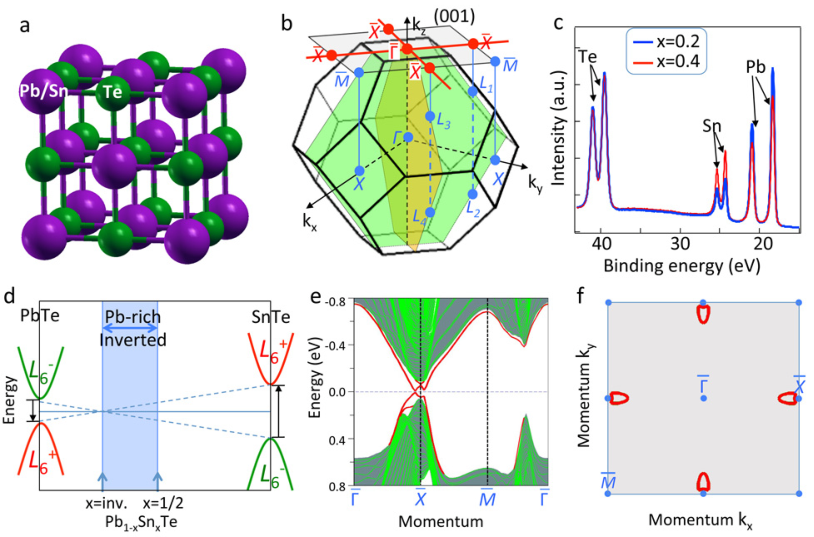}
\caption{\label{FCC} \textbf{Band inversion topological transition and surface states in Pb$_{1-x}$Sn$_{x}$Te.} \textbf{a,} The lattice of Pb$_{1-x}$Sn$_{x}$Te system is based on the sodium chloride crystal structure. \textbf{b,} The first Brillouin zone (BZ) of Pb$_{1-x}$Sn$_{x}$Te. The mirror planes are shown using green and light-brown colors. These mirror planes project onto the (001) crystal surface as the $\bar{X}-\bar{\Gamma}-\bar{X}$ mirror lines (shown by red solid lines). \textbf{c,} ARPES measured core level spectra (incident photon energy 75 eV) of two representative compositions, namely Pb$_{0.8}$Sn$_{0.2}$Te and Pb$_{0.6}$Sn$_{0.4}$Te. The photoemission core levels of tellurium $4d$, tin $4d$, and lead $5d$ orbitals are observed. \textbf{d,} The bulk band-gap of Pb$_{1-x}$Sn$_{x}$Te alloy system undergoes a band inversion upon changing the Pb/Sn ratio \cite{PST band-gap, PST Inversion1, PST Inversion2}. A TCI phase with metallic surface states is theoretically predicted when the band-gap is inverted (toward SnTe) \cite{Liang NC SnTe}.  The lowest-lying conduction and valence states associated with odd and even parity eigenvalues are labeled by $L_6^-$ and $L_6^+$, respectively. \textbf{e,f,} First-principles based calculation of band dispersion \textbf{(e)} and iso-energetic contour with energy set at 0.02 eV below the Dirac node energy \textbf{(f)} of the inverted end compound SnTe as a qualitative reference for the ARPES experiments. The surface states are shown by the red lines whereas the bulk band projections are represented by the green shaded area in \textbf{e}.}
\end{figure*}

\begin{figure}[b!]
\includegraphics[width=18cm]{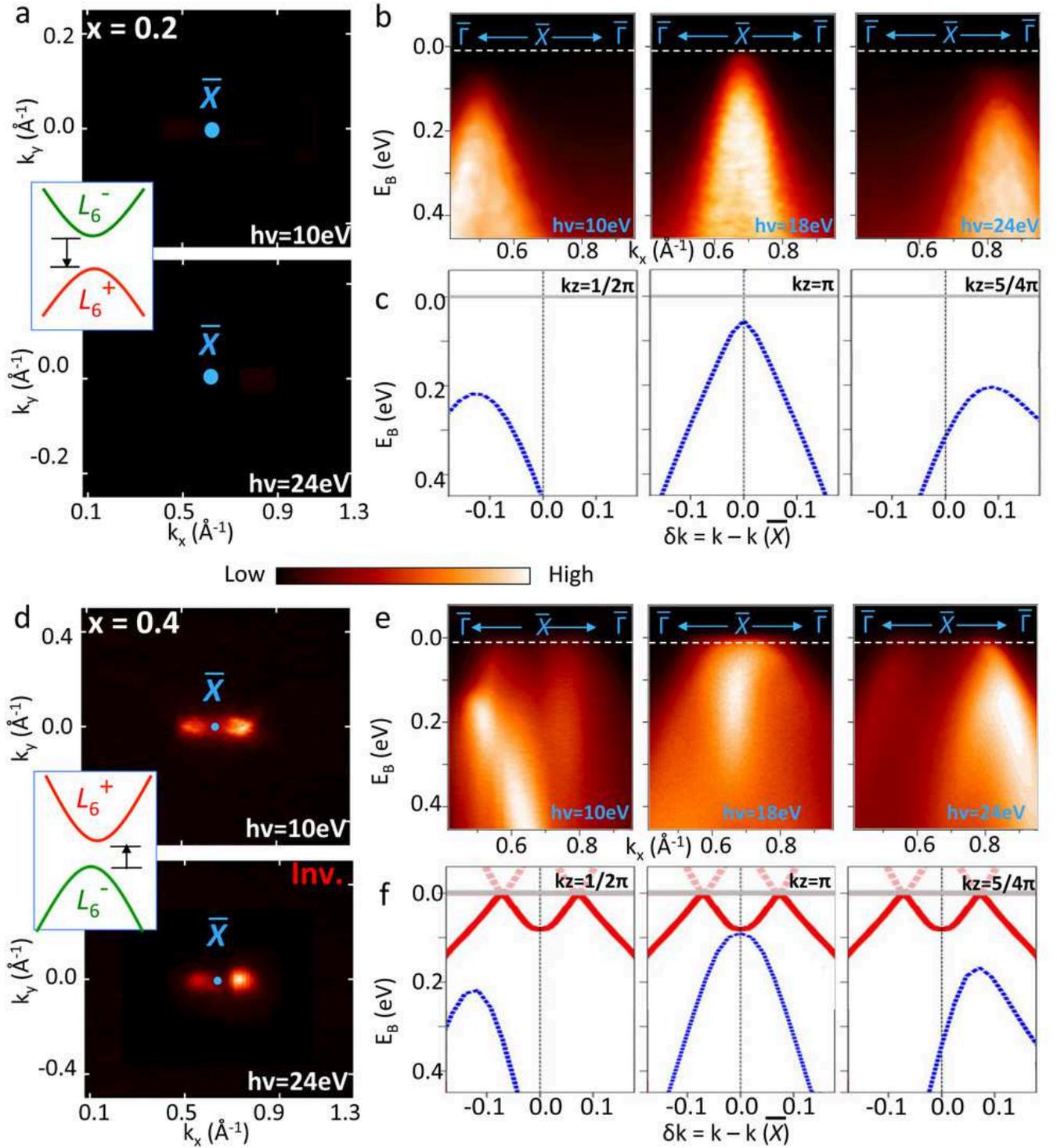}
\caption{\label{TPT} \textbf{Comparison of band insulator and topological crystalline insulator phases.}  \textbf{a, b,} ARPES low energy electronic structure measurements on Pb$_{0.8}$Sn$_{0.2}$Te ($x=0.2$). \textbf{c,} First-principle calculated electronic structure of PbTe ($x=0$) as a qualitative guide to the ARPES measurements on Pb$_{0.8}$Sn$_{0.2}$Te. PbTe, which lies on the non-inverted regime, is theoretically expected to be a conventional band-insulator \cite{Liang NC SnTe}. \textbf{d, e,} ARPES low energy electronic structure measurements on}
\end{figure}

In order to capture the electronic structure both below and above the band inversion transition (theoretically predicted to be around $x\simeq1/3$ \cite{PST Inversion1, PST Inversion2}), we choose two representative compositions, namely $x=0.2$ and $x=0.4$, for detailed systematic studies. Figure~\ref{FCC}c shows the momentum-integrated core level photoemission spectra for both compositions, where intensity peaks corresponding to tellurium $4d$, tin $4d$, and lead $5d$ orbitals are observed. The energy splitting of the Pb orbital is observed to be larger than that of the Sn orbital, which is consistent with the stronger spin-orbit coupling of the heavier Pb nuclei. In addition, the spectral intensity contribution of the Sn peaks in the $x=0.4$ sample (red) is found to be higher than that of in the $x=0.2$ sample (blue), which is also consistent with the larger Sn concentration in the $x=0.4$ samples. We perform systematic low energy electronic structure studies on these two representative compositions. Since the low energy physics of the system is dominated by the band inversion at $L$ points ($L$ points projected onto $\bar{X}$ points on the (001) surface), we present ARPES measurements with the momentum space window centered at the $\bar{X}$ point, which is the midpoint of the surface BZ edge (see Fig.~\ref{FCC}b). Figure~\ref{TPT}a,b show the ARPES Fermi surface and dispersion mappings of the Pb$_{0.8}$Sn$_{0.2}$Te sample ($x=0.2$). The system at $x=0.2$ is observed to be gapped: No band is observed to cross the Fermi level in the Fermi surface maps (Fig.~\ref{TPT}a). The dispersion measurements (Fig.~\ref{TPT}b) reveal a single hole-like band below the Fermi level. This single hole-like band is observed to show strong dependence with respect to the incident photon energy (see Fig.~\ref{TPT}b, and also detailed in Supplementary Figures S1-S3 and Supplementary Methods in the supplementary information (SI).), which reflects its three-dimensionally dispersive bulk valence band origin. As a qualitative guide to the ARPES measurements on $x=0.2$, we present first-principles based electronic structure calculation on the non-inverted end compound PbTe (Fig.~\ref{TPT}c). Our calculations confirm that PbTe is a conventional band insulator, whose electronic structure can be described as a single hole-like bulk valence band in the vicinity of each $\bar{X}$ point, which is consistent with our ARPES results on Pb-rich Pb$_{0.8}$Sn$_{0.2}$Te. The three-dimensional nature of the calculated bulk valence band is revealed by its $k_z$ evolution in Figure~\ref{TPT}c, which is in qualitative agreement with the incident photon energy dependence of our ARPES measurements shown in Figure~\ref{TPT}b.

Now we present comparative ARPES measurements under identical experimental conditions and setups on the Pb$_{0.6}$Sn$_{0.4}$Te ($x=0.4$) sample. In contrast to the conventional band insulator (insulating) behavior in the $x=0.2$ sample, the Fermi surface mapping (Fig.~\ref{TPT}d) on the $x=0.4$ sample shows two unconnected metallic Fermi pockets (dots) on the opposite sides of the $\bar{X}$ point. Such two-pockets Fermi surface topology cannot be easily interpreted as the bulk valence band since the low energy bulk valence band of the Pb$_{1-x}$Sn$_{x}$Te system is a \text{single} hole-like band, unless certain surface-umklapp processes of the bulk states are considered, which usually have only weak cross-section in ARPES measurements. The assignment of the observed two-Fermi-pockets as surface-umklapp processes or other bulk-related origins can be further ruled out by incident photon energy dependence ($k_z$ dispersion) measurements and our spin polarization studies of these metallic states (see Supplementary Figures S1-S5 in the SI). The dispersion measurements on the $x=0.4$ sample are shown in Figure~\ref{TPT}e. The single hole-like bulk valence band, which is similar to that in the $x=0.2$ sample, is also observed below the Fermi level. More importantly, a pair of metallic states crossing the Fermi level on the opposite sides of the $\bar{X}$ point is observed along the $\bar{\Gamma}-\bar{X}-\bar{\Gamma}$ mirror line momentum space direction. These states are found to show no observable dispersion upon varying the incident photon energy (further details in Supplementary Figure S3), reflecting its two-dimensional character. On the other hand, the single hole-like band is observed to disperse strongly upon varying the incident photon energy, suggesting its three-dimensional character. At a set of different photon energy ($k_z$) values, the bulk valence band intensity overlaps (intermixing) with different parts of the surface states in energy and momentum space. At a photon energy of 18 eV, the intermixing (intensity overlap) is strong, and the inner two branches of the surface states are masked by the bulk intensity. At photon energies of 10eV and 24 eV, the surface states are found to be relatively better isolated. These ARPES measurements suggest that the $x=0.4$ sample lie on the inverted composition regime and that the observed surface states are related to the band inversion transition in Pb$_{1-x}$Sn$_{x}$Te as predicted theoretically \cite{PST Inversion3, Volkov, Fradkin, Liang NC SnTe}. As a qualitative guide, we present first-principles based electronic structure calculation on the inverted end compound SnTe (Fig.~\ref{TPT}f). The calculated electronic structure of SnTe is found to be a superposition of two $k_z$ nondispersive metallic surface states and a single hole-like $k_z$ dispersive bulk valence band in the vicinity of the $\bar{X}$ point, which is in qualitative agreement with the ARPES results on Pb$_{0.6}$Sn$_{0.4}$Te. 

\bigskip
\bigskip
\textbf{Surface state topology of Pb$_{0.6}$Sn$_{0.4}$Te}
\newline
We now perform systematic measurements on the surface electronic structure of the Pb$_{0.6}$Sn$_{0.4}$Te. Figure~\ref{Basic_SS}b shows the wide range Fermi surface mapping covering the first surface BZ. The surface states are observed to be present, and only present, along the mirror line ($\bar{\Gamma}-\bar{X}-\bar{\Gamma}$) directions. No other states are found along any other momentum directions on the Fermi level. In close vicinity to each $\bar{X}$ point, a pair of surface states are observed along the mirror line direction. One lies inside the first surface BZ while the other is located outside. Therefore, in total four surface states are observed within the first surface BZ, in agreement with the fact that there are four band inversions in Pb$_{1-x}$Sn$_{x}$Te. The mapping zoomed-in near the $\bar{X}$ point (Fig.~\ref{Basic_SS}c) reveals two unconnected small pockets (dots). The momentum space distance from the center of each pocket to the $\bar{X}$ point is about $0.09$ $\textrm{\AA}^{-1}$. Dispersion measurements ($E_\textrm{B}$ vs $k_{\/}$) are performed along three important momentum space cuts, namely cuts 1, 2, and 3 defined in Fig.~\ref{Basic_SS}b, in order to further reveal the electronic structure of the surface states. Metallic surface states crossing the Fermi level are observed in both cuts 1 and 2, whereas cut 3 is found to be fully gapped, which is consistent with the theoretically calculated surface states electronic structure shown in Figure~\ref{FCC}e. In cut 1 (Fig.~\ref{Basic_SS}d), which is the mirror line ($\bar{\Gamma}-\bar{X}-\bar{\Gamma}$) direction, a pair of surface states are observed on the Fermi level. The surface states in our Pb$_{0.6}$Sn$_{0.4}$Te samples are found to have a relatively broad spectrum, which can be possibly understood by the strong scattering in the disordered alloy system similar to the broad spectrum of the topological surface states in the Bi$_{1-x}$Sb$_x$ alloy \cite{David Nature BiSb}. In addition to the scattering broadening, the surface states are also observed to tail on the very strong main valence band emission (for example, the white region of intensity distribution in Fig.~\ref{Basic_SS}d). In the case of Figure~\ref{TPT}e (photon energy of 24 eV), the bulk valence band maximum (VBM) locates outside the first surface BZ. The surface state inside the first surface BZ is relatively better isolated from the bulk bands as compared to the one outside the first surface BZ. We thus study the dispersion along cut 2 (Fig.~\ref{Basic_SS}d), which only cuts across the surface states inside the first surface BZ. Both the dispersion maps and the momentum distribution curves in cut 2 reveal that the surface states along cut 2 are nearly Dirac-like (linearly dispersive) close to the Fermi level. In many topological insulators, the surface states deviate from ideal linearity \cite{Liang PRL Warping}. Fitting of the momentum distribution curves of cut 2 (see Supplementary Figure S6 for data analysis) shows that the experimental chemical potential ($E_\textrm{F}$) lies roughly at (or just below) the Dirac point energy ($E_\textrm{D}$), $E_\textrm{F}=E_\textrm{D}\pm0.02$ eV. The surface states' velocity is obtained to be $2.8{\pm}0.1$ $\textrm{eV}{\cdot}\textrm{\AA}$ ($4.2\pm0.2$ ${\times}10^5$ m/s) along cut 2, and $1.1{\pm}0.3$ $\textrm{eV}{\cdot}\textrm{\AA}$ ($1.7\pm0.4$ $\times10^5$ m/s) for the two outer branches along cut 1, respectively. 

\bigskip
\bigskip
\textbf{Spin polarization measurements of the surface states}
\newline

We study the spin polarization of the low energy states of the Pb$_{0.6}$Sn$_{0.4}$Te samples, which are highly dominated by the surface states near the Fermi level. We further compare and contrast their spin behavior with that of the states at high binding energies away from the Fermi level, which are highly dominated by the bulk valence band in the same sample. Spin-resolved (SR) measurements are performed in the spin-resolved momentum distribution curve mode \cite{Spin1, Spin2}, which measures the spin-resolved intensity and net spin polarization at a fixed binding energy along a certain momentum space cut direction (detailed in the Methods section). As shown in Figure~\ref{Spin}a, our spin-resolved measurements are performed along the mirror line ($\bar{\Gamma}-\bar{X}-\bar{\Gamma}$) direction, since the electronic and spin structure along this direction is most critically relevant to the predicted TCI phase \cite{Liang NC SnTe}. Considering that the natural Fermi level of our $x=0.4$ samples are very close to the Dirac point (which is spin degenerate), the spin polarization of the surface states are measured at 60 meV below the Fermi level in order to gain proper contrast, namely SR-Cut 1 in Figure~\ref{Spin}a. As shown in the net spin polarization measurement of SR-Cut 1 in Figure~\ref{Spin}c, in total four spins pointing in the ($\pm$) in-plane tangential direction are revealed for the surface states along the mirror line direction. This is well consistent with the observed two surface state cones (four branches in total) near an $\bar{X}$ point along the mirror line direction. To compare and contrast the spin polarization behavior of the surface states (SR-Cut 1) with that of the bulk states, we perform spin-resolved measurement SR-Cut 2 at $E_\textrm{B}=0.70$ eV, where the bulk valence bands are prominently dominated. Indeed, in contrast to SR-Cut 1 reflecting the surface states' spin polarization, no significant net spin polarization is observed for SR-Cut2, which is expected for the bulk valence bands of the inversion symmetric (centrosymmetric) Pb$_{1-x}$Sn$_{x}$Te system. Our observed spin polarization configuration of the surface states is also in qualitative agreement with the first-principles calculation spin texture of the SnTe TCI surface states (see Supplementary Figures S7-S9 for texture calculation). And Experimental derived topological invariant (Mirror Chern number \cite{Mirror Chern Number, David Science BiSb}) $n_{\textrm{M}}=-2$ also agrees with theoretical prediction for SnTe \cite{Liang NC SnTe}. 

\bigskip
\bigskip
\textbf{Discussion}
\newline
We discuss the possibility that the observed surface states in our data are the signature of the theoretically predicted TCI phase. The TCI phase was predicted to be observed in the band inverted side and absent in the non-inverted side \cite{Liang NC SnTe}. In our data of the Pb$_{0.6}$Sn$_{0.4}$Te samples which lie on the inverted side, surface states are observed, and only observed, along the two independent mirror line ($\bar{\Gamma}-\bar{X}-\bar{\Gamma}$) directions. Within the first surface BZ, two surface states are observed on each mirror line, which are found to locate in vicinity of the $\bar{X}$ points. Dispersion measurements (e.g. Cut 2 in Fig.~\ref{Basic_SS}) reveal a nearly linear dispersion. Spin polarization measurements show that these surface states are spin-polarized and their spin polarization direction is locked with their momentum (spin-momentum locking). The overall electronic structure observed for our surface states are in qualitative agreement with the theoretically predicted surface states of the TCI phase in SnTe \cite{Liang NC SnTe}. It is important to note that in an alloy, the space group symmetry is only of the averaged type. The spatial mirror symmetry required for the topological protection is only preserved in a globally averaged sense. To this date, it is not even theoretically known whether only averaged mirror symmetry is sufficient for the predicted TCI phase. In the absence of any theoretical work for the alloy, the experimental proof of mirror protection in the alloy Pb$_{1-x}$Sn$_{x}$Te(Se) system perhaps requires the observation of strictly gapless surface states. The current ARPES works \cite{PSS TCI, PST Xu} have not demonstrated resolution better than 5 meV, and thus fitting of the energy-momentum distribution curves lacks the resolving power to exclude gap values less than 5 meV (demonstrated in Supplementary Figure S10 and Supplementary Discussion). On the other hand, a proof of strictly gapless surface states in fact requires a surface and spin sensitive experimental setup with energy resolution better than 1 meV, which can even exclude possible small gap values of $\leq1$ meV at the Dirac point. Even a gap size of even less than 1 meV can lead to observable change in spin-transport experiments, thus destroying the delicate TCI phase. Therefore, while the existence of surface states correlated with band inversion transition is established here in our work, their gapped or gaplessness nature requires surface transport experiments on samples with improved quality, which is currently not possible \cite{PSS TCI}. 

Our observations of spin-momentum locked surface states in the inverted composition and their absence in the non-inverted composition suggest that Pb$_{1-x}$Sn$_{x}$Te system is a tunable spin-orbit insulator analogous to the BiTl(S$_{1-{\delta}}$Se$_{\delta}$)$_{2}$ system \cite{Hasan QPT}, which features a topological phase transition across the band inversion transition. However, the key difference is that odd number of band inversions in the BiTl(S$_{1-{\delta}}$Se$_{\delta}$)$_{2}$ system leads to odd number of surface states, whereas even number of band inversions in Pb$_{1-x}$Sn$_{x}$Te system leads to even number of surface states per surface BZ. We further present a comparison of the Pb$_{0.6}$Sn$_{0.4}$Te and a single Dirac cone Z$_2$ topological insulator (TI) GeBi$_2$Te$_4$ \cite{Ternary arXiv, Ternary PRB, Kimura}. As shown in Figure~\ref{TCITI}a-c, for GeBi$_2$Te$_4$, a single surface Dirac cone is observed enclosing the time-reversal invariant (Kramers') momentum $\bar{\Gamma}$ in both ARPES and calculation results, demonstrating its Z$_2$ topological insulator state and the time-reversal symmetry protection of its single Dirac cone surface states. On the other hand, for the Pb$_{0.6}$Sn$_{0.4}$Te samples (Fig.~\ref{TCITI}d-f), none of the surface states is observed to enclose any of the time-reversal invariant momentum, suggesting their irrelevance to the time-reversal symmetry related protection. With future ultra-high-resolution experimental studies to prove the strict gapless nature of the Pb$_{0.6}$Sn$_{0.4}$Te surface states and therefore the topological protection by the crystalline mirror symmetries, it is then possible to realize magnetic yet topologically protected surface states in the Pb$_{1-x}$Sn$_{x}$Te system due to its irrelevance to the time-reversal symmetry related protection, which is fundamentally not possible in the Z$_2$ topological insulator systems. Considering the wide applications of magnetic materials in modern electronics, such topologically protected surface states compatible with magnetism will be of great interest in terms of integrating topological insulator materials into future electronic devices.

We note that one advantage of the Pb$_{1-x}$Sn$_{x}$Te system is that it can be easily doped with manganese, thallium, or indium to achieve bulk magnetic or superconducting states \cite{PbMnTe, Tl-PbTe ARPES, In-SnTe Ando}. The symmetry in the Pb$_{1-x}$Sn$_{x}$Te system (its nonmagnetic character) can be broken by magnetic or superconducting doping into the bulk or the surface. In future experiments, it would be interesting to explore the modification of our observed surface states brought out by magnetic and superconducting correlations, in order to search for exotic magnetic and superconducting order on the surface. Such magnetic and superconducting orders on the surface states in Pb$_{1-x}$Sn$_{x}$Te can be different from those recently observed in the Z$_2$ topological insulators \cite{Andrew CuBiSe, Hedgehog} due to its very distinct topology of surface electronic structure. The novel magnetic and superconducting states to be realized with this novel topology are not strongly related to the question of gapless or gapped nature of the TCI phase. Therefore our observation of the spin-polarized surface states presented here provides the much desired platform for realizing unusual surface magnetic and superconducting states in future experiments.

\bigskip
\bigskip
\textbf{Methods}
\newline
\textbf{Electronic structure measurements.}
Spin-integrated angle-resolved photoemission spectroscopy (ARPES) measurements were performed with incident photon energies of $8$ eV to $30$ eV at beamline 5-4 at the Stanford Synchrotron Radiation Lightsource (SSRL) in the Stanford Linear Accelerator Center, and with $28$ eV to $90$ eV photon energies at beamlines 4.0.3, 10.0.1, and 12.0.1 at the Advance Light Source (ALS) in the Lawrence Berkeley National Laboratory (LBNL). Samples were cleaved \textit{in situ} between $10$ to $20$ K at chamber pressure better than $5 {\times} 10^{-11}$ torr at both the SSRL and the ALS, resulting in shiny surfaces. Energy resolution was better 15 meV than and 1\% of the surface BZ. In order to cross-check the cleavage surface orientation in ARPES measurements, the Laue back-reflection x-ray measurements were performed at the ALS using a Realtime Laue delay-line detector from Multiwire Laboratories (Ithaca, NY) and diffraction spot indexing was performed using the MWL NorthStar orientation package. The x-ray Laue measurements along with the analysis results are presented in the Supplementary Figure S11.

\textbf{Spin-resolved measurements.}
Spin-resolved ARPES measurements were performed on the SIS beamline at the Swiss Light Source (SLS) using the COPHEE spectrometer with two 40kV classical Mott detectors and a photon energy of 24 eV, which systematically measures all three components of the spin of the electron ($P_x$, $P_y$, and $P_z$) as a function of its energy and momentum \cite{Spin1, Spin2}. Energy resolution was better than 60 meV and 3\% of the surface BZ for the spin-resolved measurements. Samples were cleaved \textit{in situ} at 20 K at chamber pressure less than $2{\times}10^{-10}$ torr. Typical electron counts on the detector reach $5 \times 10^5$, which places an error bar of approximately $\pm0.01$ for each point on our measured polarization curves.

\textbf{Sample growth.}
Single crystals of Pb$_{1-x}$Sn$_{x}$Te were grown by the standard growth method (described in Ref. \cite{Growth} for the end compound PbTe). High purity (99.9999\%) elements of lead, tin, and tellurium were mixed in the right ratio. Lead was etched by CH$_3$COOH+H$_2$O$_2$ (4:1) to remove the surface oxide layer before mixing. The mixture was heated in a clean evacuated quartz tube to 924 $^{\circ}\textrm{C}$ where it was held for two days. Afterwards, it was cooled slowly, at a rate of $1.5$ $^{\circ}\textrm{C}{\cdot}h^{-1}$ in the vicinity of the melting point, from the high temperature zone towards room temperature. The nominal concentrations were estimated by the Pb/Sn mixture weight ratio before the growth, in which the two representative compositions presented in the paper were estimated to be $x_{\textrm{nom}}=0.20$ and $x_{\textrm{nom}}=0.36$. The concentration values were re-examined on the single crystal samples after growth using high resolution x-ray diffraction measurements by tracking the change of the diffraction angle (the 2Theta angle) of the sharpest Bragg peak, from which we obtained $x=0.22$ and $x=0.39$ (see Supplementary Figure S12 for diffraction data). In the paper we note the chemical compositions as Pb$_{0.8}$Sn$_{0.2}$Te and Pb$_{0.6}$Sn$_{0.4}$Te.

\textbf{First-principles calculation methods.}
We perform first-principles calculations to extract both the electronic structure and the spin texture of the SnTe surface states. Our first-principles calculations are within the framework of the density functional theory (DFT) using full-potential projected augmented wave method \cite{PAW} as implemented in the VASP package \cite{PBE}. The generalized gradient approximation (GGA) \cite{VASP} is used to model exchange-correlation effects. The spin-orbital coupling (SOC) is included in the self-consistent cycles. The surfaces are modeled by periodically repeated slabs of 33-atomic-layer thickness, separated by 13-angstrom-wide vacuum regions, using a $12\times12\times1$ Monkhorst-Pack $k$-point mesh over the Brillouin zone (BZ) with 208 eV cutoff energy. The room-temperature crystal structures of SnTe in ideal Òsodium chlorideÓ are used to construct the slab in order to have the required crystal structure and mirror symmetries required for the predicted TCI phase \cite {Liang NC SnTe} (The rhombohedral distortion is found to only occur in the Sn-rich compositions ($x>0.5$) and only at low temperature \cite{PST Rhombohedral Distortion}). The experimental lattice constants of SnTe with the value of 6.327 $\textrm{\AA}$ are used \cite{LSalts}. The experimental lattice constant of PbTe with the value of 6.46 $\textrm{\AA}$ \cite{LSalts} in ideal sodium chloride lattice are used in the PbTe band structure calculation.

\*Correspondence and requests for materials should be addressed should be addressed to M.Z.H. (Email: mzhasan@princeton.edu).

\newpage

\addtocounter{figure}{-1}
\begin{figure} [t!]
\caption{(Previous page.) Pb$_{0.6}$Sn$_{0.4}$Te ($x=0.4$) under identical experimental conditions and setups as in \textbf{(a and b)}. \textbf{f,} First-principle calculated electronic structure of SnTe ($x=1$) as a qualitative guide to the ARPES measurements on Pb$_{0.6}$Sn$_{0.4}$Te. SnTe, which lies on the inverted regime, is theoretically expected to be a topological crystalline insulater \cite{Liang NC SnTe}. The insets of panels \textbf{a and d} show the non-inverted and inverted bulk band-gap of the $x=0.2$ and $x=0.4$ samples, respectively. The blue and red lines in panels \textbf{c and f} are the calculated dispersion of the bulk bands and the surface states, respectively.}
\end{figure}

\newpage
\begin{figure}
\centering
\includegraphics[width=18cm]{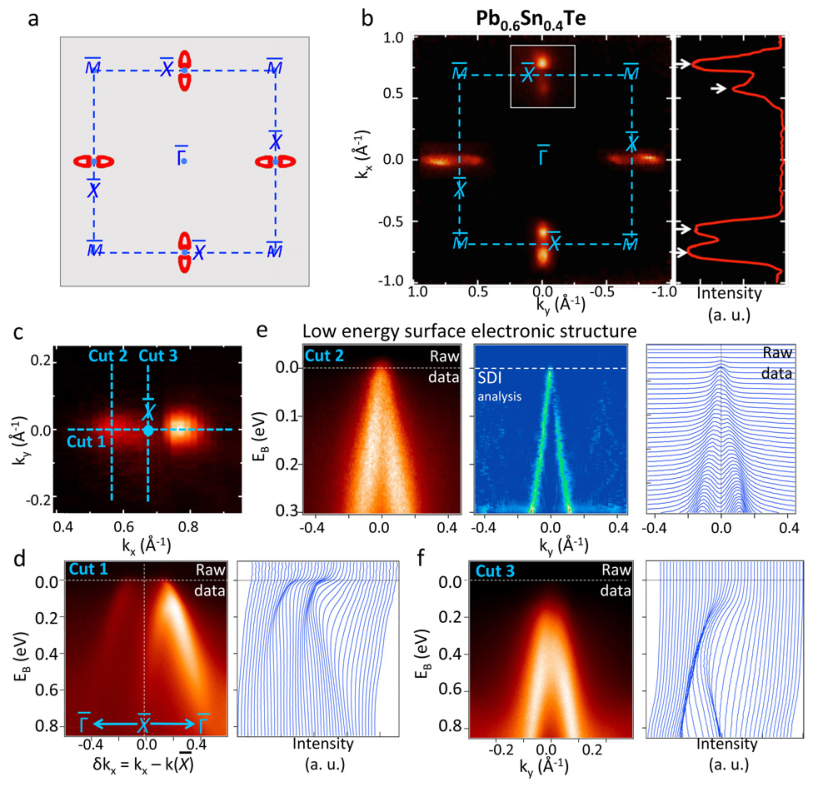}
\caption{\label{Basic_SS} \textbf{Dirac surface states and topological crystalline insulator phase in Pb$_{0.6}$Sn$_{0.4}$Te.} \textbf{a,} First-principle calculated surface states of SnTe with energy set at 0.02 eV below the Dirac node energy are shown in red. The blue dotted lines mark the first surface BZ. \textbf{b,} Left panel: ARPES iso-energetic contour mapping ($E_\textrm{B}=0.02$ eV) of Pb$_{0.6}$Sn$_{0.4}$Te covering the first surface Brillouin zone (BZ) using incident photon energy of 24 eV (throughout Fig. 3). Right panel: Spectral intensity distribution as a function of momentum along the horizontal mirror line (defined by $k_y=0$). \textbf{c,} High resolution Fermi surface mapping ($E_\textrm{B}=0.0$ eV) in the vicinity of one of the $\bar{X}$ points (indicated by the white square in \textbf{b}). \textbf{d-f,} Dispersion maps ($E_\textrm{B}$ vs $k_{\/}$) and corresponding energy (momentum) distribution curves of the momentum space }
\end{figure}
\addtocounter{figure}{-1}
\begin{figure} [t!]
\caption{(Previous page.) cuts 1, 2, and 3. The momentum space cut-directions of cuts 1, 2, and 3 are defined by blue dotted lines in panel \textbf{c}. The second derivative image (SDI) of the measured dispersion is additionally shown for \textbf{d}.}
\end{figure}

\newpage
\begin{figure}
\centering
\includegraphics[width=18cm]{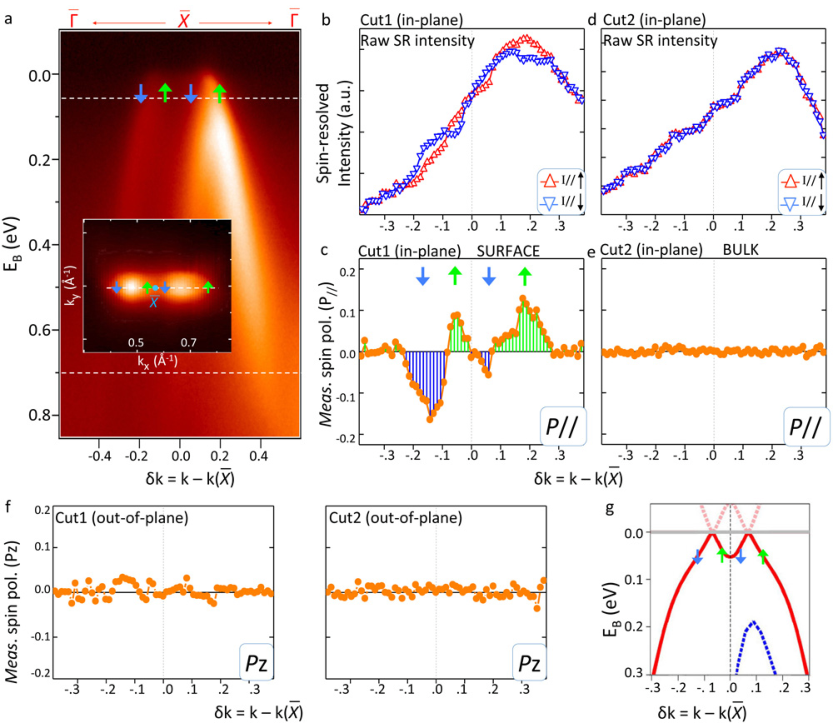}
\caption{\label{Spin} \textbf{Spin polarization of topological Dirac surface states.} \textbf{a,} ARPES dispersion map along the mirror line direction. The white dotted lines show the binding energies chosen for spin-resolved measurements, namely SR-Cut 1 at $E_\textrm{B}=0.06$ eV and SR-Cut 2 $E_\textrm{B}=0.70$ eV. Inset: Measured spin polarization profile is shown by the green and blue arrows on top of the ARPES iso-energetic contour at binding energy $E_\textrm{B}=0.06$ eV for SR-Cut1. \textbf{b,c,} Measured in-plane spin-resolved intensity \textbf{(b)} and in-plane spin polarization \textbf{(c)} of the surface states (SR-Cut 1) near the Fermi level at $E_\textrm{B}=0.06$ eV. \textbf{d,e} Measured in-plane spin-resolved intensity \textbf{(d)} and in-plane spin polarization \textbf{(e)} of the bulk valence bands (SR-Cut 2) at $E_\textrm{B}=0.70$ eV. \textbf{f,} Out-of-plane spin polarization measurements of SR-Cut 1 and SR-Cut 2. The error bar is $\pm0.01$ for data points in all spin polarization measurements. \textbf{g,} Theoretically expected spin polarization configuration of the surface states which corresponds to a mirror topological invariant}
\end{figure}
\addtocounter{figure}{-1}
\begin{figure} [t!]
\caption{(Previous page.)  (the mirror Chern number) $n_{\textrm{M}}=-2$ \cite{Mirror Chern Number, David Science BiSb}. Measured spin polarization texture configuration (green and blue arrows) of the surface states (SR-Cut 1) is shown on top of the calculated surface states. The blue and red lines are the calculated bulk bands and surface states respectively.}
\end{figure}

\newpage
\begin{figure}
\centering
\includegraphics[width=18cm]{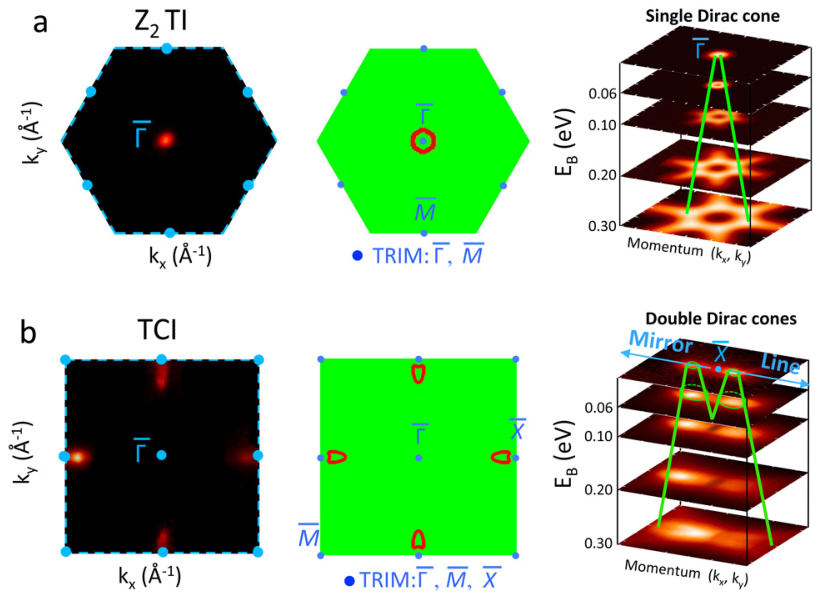}
\caption{\label{TCITI} \textbf{A comparison of Z$_2$ (Kane-Mele) topological insulator and topological crystalline insulator (TCI) phases.} \textbf{a,} ARPES and calculation results of the surface states of a Z$_2$ topological insulator GeBi$_2$Te$_4$, an analog to Bi$_2$Se$_3$ \cite{Matthew Nature physics BiSe}. Left: ARPES measured Fermi surface with the chemical potential tuned near the surface Dirac point. Middle: First-principles calculated iso-energetic contour of the surface states near the Dirac point. Right: A stack of ARPES iso-energetic contours near the $\bar{\Gamma}$ point of the surface BZ. \textbf{b,}  ARPES measurements on the Pb$_{0.6}$Sn$_{0.4}$Te ($x=0.4$) samples and band calculation results on the end compound SnTe. The green straight lines in both \textbf{(a) and (b)} represent the linear energy-momentum dispersion of the Dirac cones. The green circles in \textbf{(b)} represent the double Dirac cone contours near each $\bar{X}$ point on the surface of Pb$_{0.6}$Sn$_{0.4}$Te.}
\end{figure}

\setcounter{figure}{0}

\renewcommand{\figurename}{\textbf{Supplementary Figure}}

\clearpage
\textbf{
\begin{center}
{\large \underline{Supplementary Information}: \\Observation of a topological crystalline insulator phase and topological phase transition in Pb$_{1-x}$Sn$_x$Te}
\end{center}
}

\vspace{0.2cm}

\begin{center}
Su-Yang Xu, Chang Liu, N. Alidoust, M. Neupane, D. Qian, I. Belopolski, J. D. \\
Denlinger, Y. J. Wang, H. Lin, L. A. Wray, G. Landolt, B. Slomski, J. H. Dil,  A. \\
Marcinkova, E. Morosan, Q. Gibson, R. Sankar, F. C. Chou, R. J. Cava,\\
A. Bansil, and M. Z. Hasan
\end{center}

\vspace{0.25cm}

\textbf{
\begin{center}
{\large This file includes:\\}
\end{center}
}
\vspace{0.45cm}
\begin{tabular}{l}
Supplementary Figures\\
Supplementary Discussion\\
Supplementary Methods\\
Supplementary References\\
\end{tabular}


\clearpage

\textbf{\large {Supplementary Figures}}
\bigskip
\begin{figure*}[h]
\includegraphics[width=17cm]{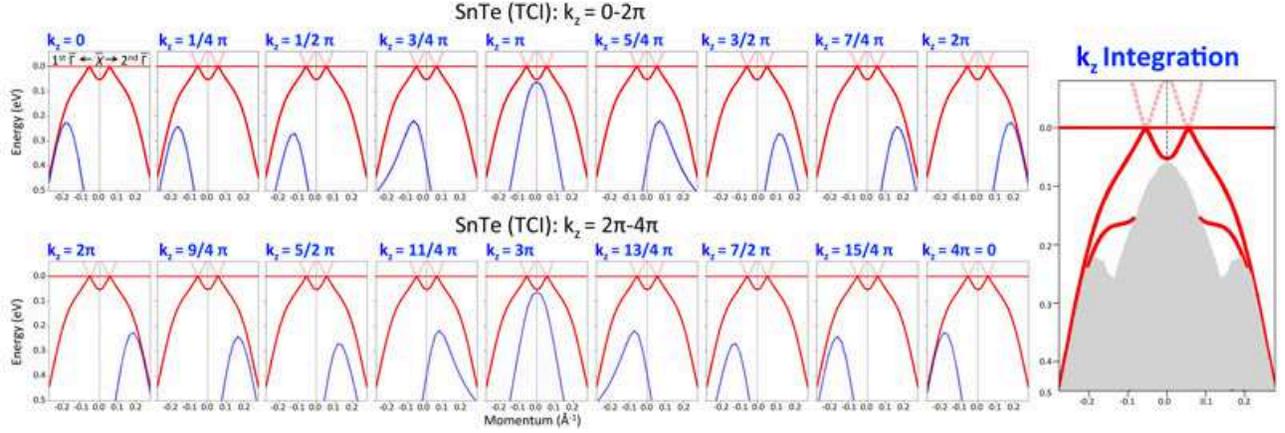}
\centering
\caption{\label{SnTe_BVB} \textbf{Calculation of bulk band dispersion in SnTe.} Theoretical calculated electronic structure of SnTe is shown along $\bar{\Gamma}$(first surface BZ)$-\bar{X}-\bar{\Gamma}$(second surface BZ) mirror line direction. The SnTe lattice in ideal sodium chloride structure without rhombohedral distortion is used in the calculation. The experimental lattice constants of SnTe with the value of 6.327 $\textrm{\AA}$ are used. Therefore the TCI surface states are realized in the calculation. The red and blue lines correspond to the TCI surface states and the bulk bands respectively. The dispersion of the surface states is found to be independent of the $k_z$ values, whereas the bulk bands are found to show strong $k_z$ dispersion with a $k_z$ period of $4\pi$. Right panel: Theoretical calculation of SnTe surface states and projection of band bands, showing how the surface states are connected to the bulk bands. The bulk band projection is the bulk valence band integrated over the entire $k_z$ range from $k_z=0$ to $k_z=4\pi$.}
\end{figure*}

\clearpage
\begin{figure*}[t]
\includegraphics[width=17cm]{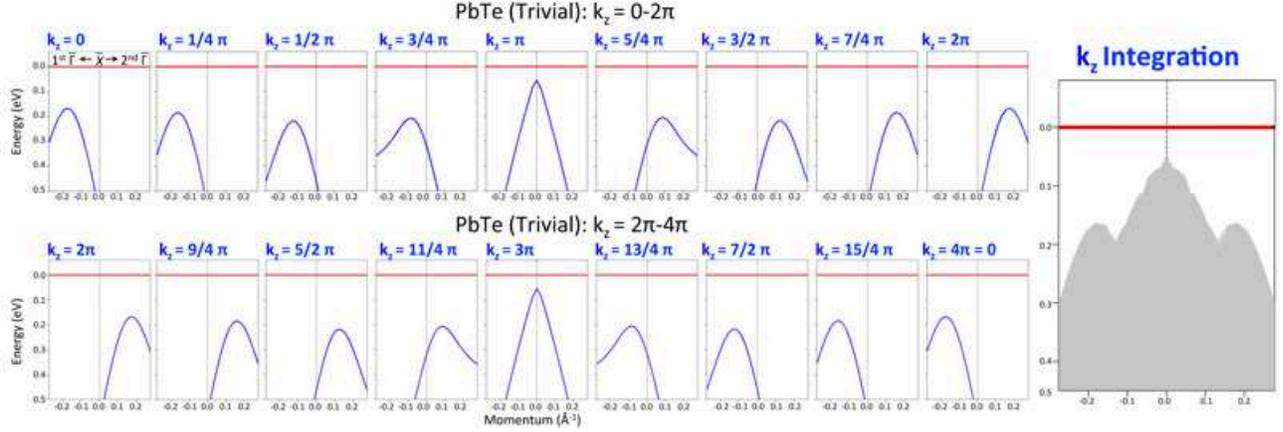}
\centering
\caption{\label{PbTe_BVB} \textbf{Calculation of bulk band dispersion in PbTe.} Theoretical calculated electronic structure of PbTe is shown along $\bar{\Gamma}$(first surface BZ)$-\bar{X}-\bar{\Gamma}$(second surface BZ) mirror line direction. The experimental lattice constant of PbTe with the value of 6.46 $\textrm{\AA}$ in ideal sodium chloride are used in the PbTe band structure calculation. The trivial insulator phase is realized in PbTe and no surface states is found in the calculation. The blue lines correspond to the bulk bands at different $k_z$ values. The bulk bands are found to show strong $k_z$ dispersion with a period of $4\pi$. Right panel: Theoretical calculation of PbTe projection of band bands. The bulk band projection is the bulk valence band integrated over the entire $k_z$ range from $k_z=0$ to $k_z=4\pi$.}
\end{figure*}

\clearpage
\begin{figure*}[h]
\includegraphics[width=17cm]{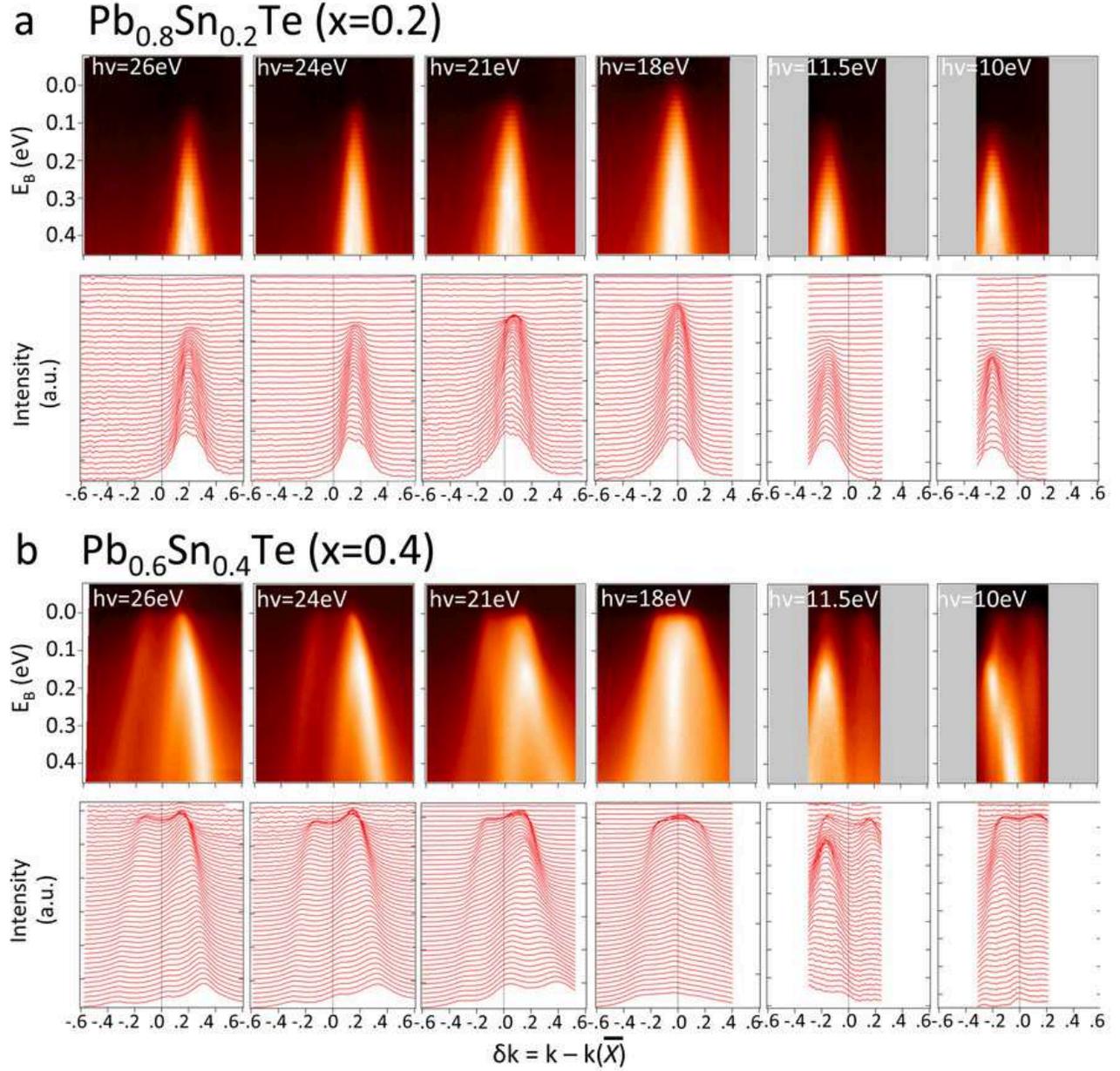}
\caption{\label{hv_dep_com} \textbf{Incident photon energy dependence study of Pb$_{0.8}$Sn$_{0.2}$Te and Pb$_{0.6}$Sn$_{0.4}$Te.} \textbf{a,} The ARPES dispersion maps and corresponding momentum distribution curves (MDCs) along the $\bar{\Gamma}$(first surface BZ)$-\bar{X}-\bar{\Gamma}$(second surface BZ) mirror line direction at different incident photon energies for the non-inverted composition Pb$_{0.8}$Sn$_{0.2}$Te. \textbf{b,} The ARPES dispersion maps and corresponding momentum distribution curves (MDCs) along the $\bar{\Gamma}-\bar{X}-\bar{\Gamma}$ mirror line direction at different incident photon energies for the inverted composition Pb$_{0.6}$Sn$_{0.4}$Te. For the non-inverted $x=0.2$ sample, no surface states on the Fermi level is observed for all the incident photon energy values applied under the same experimental conditions and setups. The $k_z$ evolution (dispersion) of the bulk valence band at different $k_z$ values (different incident photon energies) is}
\end{figure*}
\addtocounter{figure}{-1}
\begin{figure} [t!]
\caption{(Previous page.) in qualitative agreement with the theoretical calculation results shown above. For the inverted $x=0.4$ sample, a pair of metallic states is observed on the opposite sides of each $\bar{X}$ point. The dispersion of the metallic states on the Fermi level is found to show no observable change upon varying the incident photon energy. The bulk valence band is observed to be a single hole-like band below the Fermi level at each photon energy, which shows strong dispersion with respect to the incident photon energy.}
\end{figure}

\clearpage
\begin{figure*}[h]
\includegraphics[width=17cm]{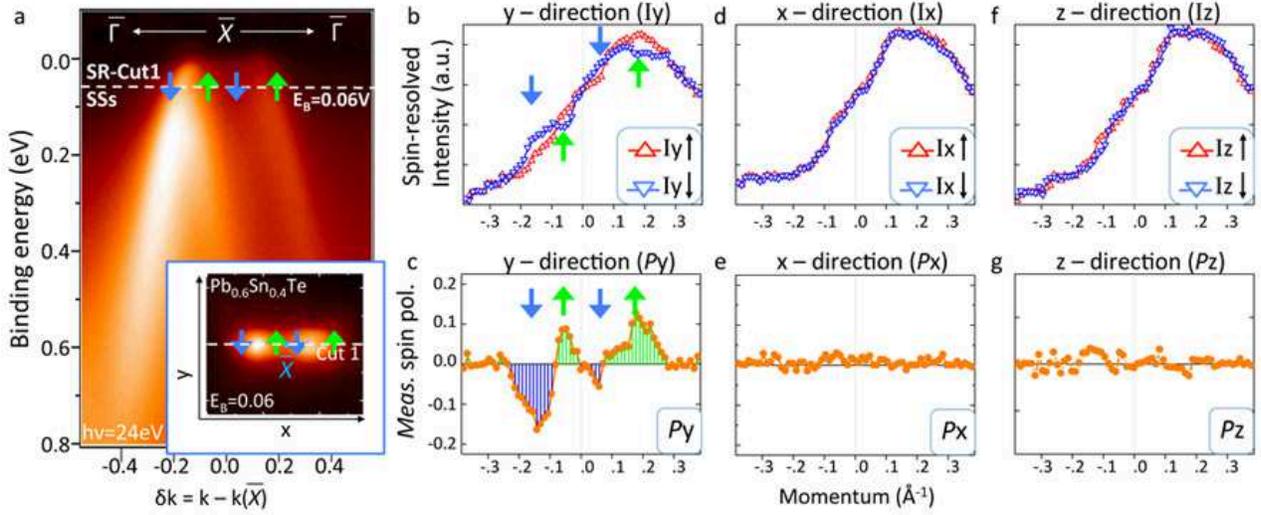}
\centering
\caption{\label{Spin_cut1} \textbf{Spin measurements near the surface state Fermi level.} \textbf{a,} ARPES measured dispersion map along the $\bar{\Gamma}-\bar{X}-\bar{\Gamma}$ mirror line direction. The white dotted line shows the binding energy chosen for the spin-resolved measurement of SR-Cut 1. SR-Cut1 is at binding energy near the Fermi level ($E_{\textrm{B}}=0.06$ eV), and thus it measures the spin polarization of the surface states.  Inset shows the momentum space direction of SR-Cut 1 in the $(k_x,k_y)$ momentum space plane on top of the measured iso-energetic contour of the surface states at $E_{\textrm{B}}=0.06$ eV. The blue and green arrows represent the measured spin polarization configuration of SR-Cut1. \textbf{b-c,} The spin-resolved intensity along $\pm\hat{y}$ direction, and  the $\hat{y}$ component of the measured spin polarization of SR-Cut 1. In total four spin vectors pointing along the $\pm\hat{y}$ direction are clearly revealed by the spin-resolved measurements. \textbf{d-e, and f-g,} Same as \textbf{b,c,} but for the $\hat{x}$ and $\hat{z}$ components of the spin, respectively. No significant $\hat{x}$ and $\hat{z}$ components of spin polarization is observed.}
\end{figure*}

\clearpage
\begin{figure*}[h]
\includegraphics[width=17cm]{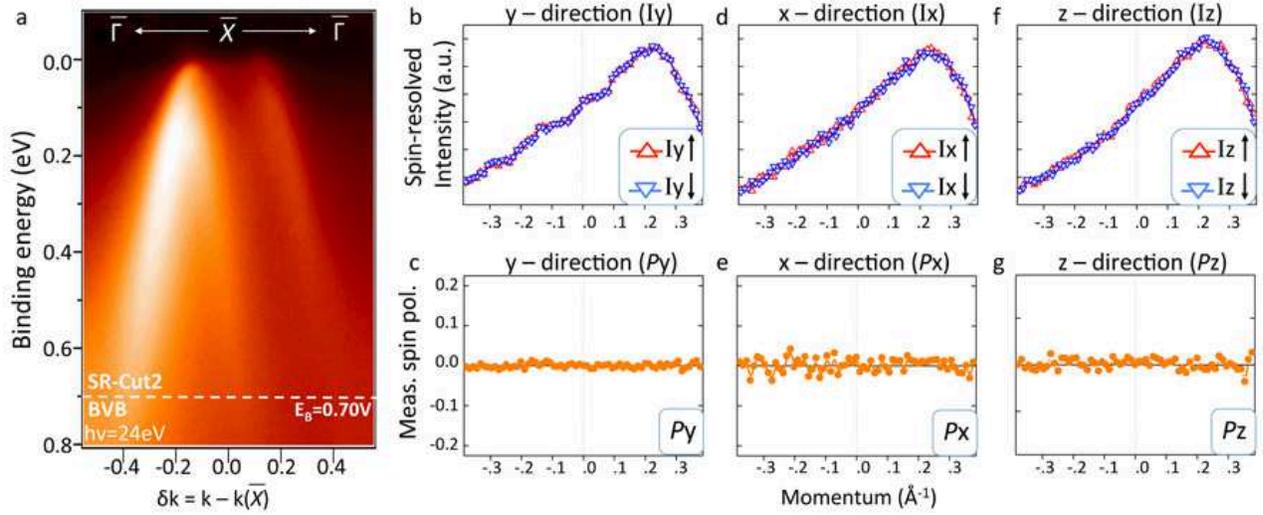}
\centering
\caption{\label{Spin_cut2} \textbf{Spin measurements at high binding energy (bulk valence bands).} \textbf{a,} ARPES measured dispersion map along the $\bar{\Gamma}-\bar{X}-\bar{\Gamma}$ mirror line direction. The white dotted line shows the binding energy chosen for the spin-resolved measurement of SR-Cut 2. SR-Cut 2 is at high binding energy far away from the Fermi level ($E_{\textrm{B}}=0.70$ eV), and thus it measures the spin polarization of the bulk valence bands. \textbf{b-c, d-e, and f-g} The spin-resolved intensity and the measured spin polarization of SR-Cut 2 along the $\hat{y}$, $\hat{x}$, and $\hat{z}$ directions, respectively. No significant spin polarization is observed for SR-Cut 2 (the bulk valence bands).}
\end{figure*}

\clearpage
\begin{figure*}[h]
\includegraphics[width=17cm]{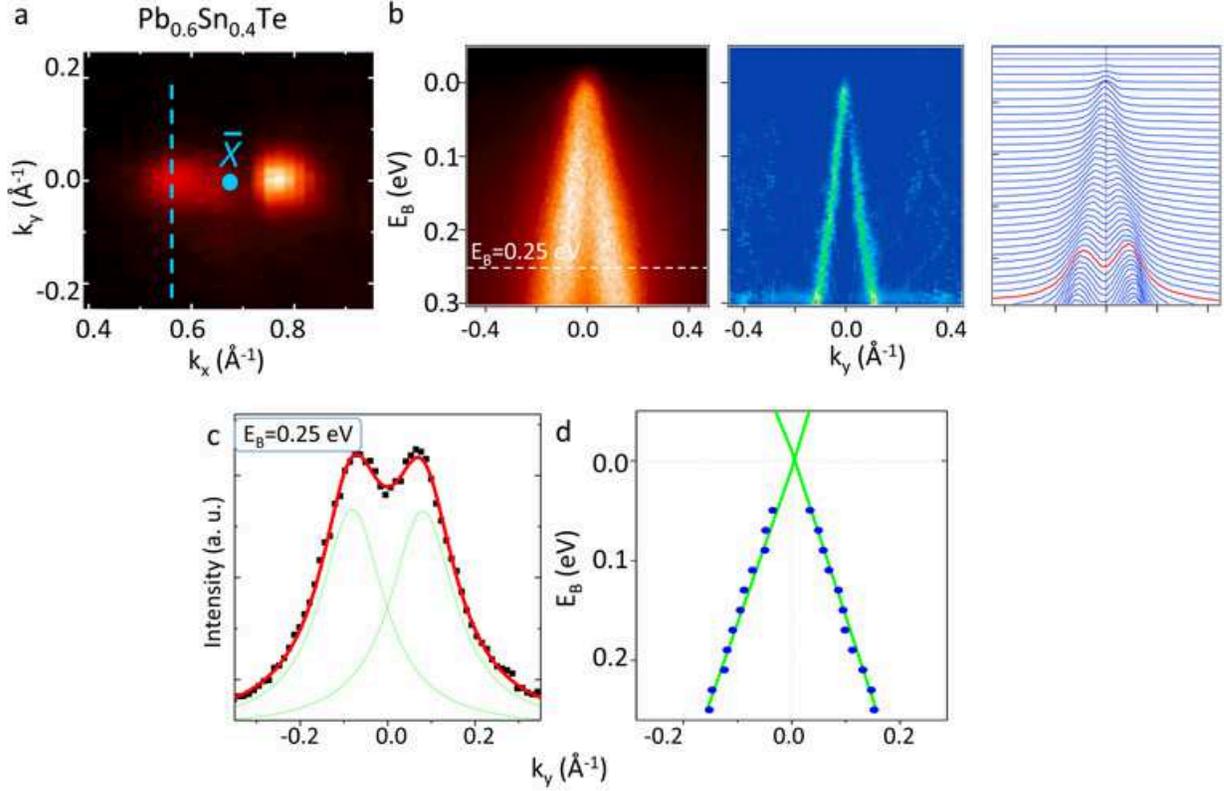}
\caption{\label{Fitting} \textbf{Surface state lineshape.} \textbf{a,} Fermi surface mapping zoomed-in in the vicinity of an $\bar{X}$ point. \textbf{b,} Dispersion map, second derivative analysis, and the corresponding momentum distribution curves of the surface states along momentum space cut direction defined by the blue dotted lines in \textbf{a}. The momentum distribution curve at $E_{\textrm{B}}=0.25$ eV is highlighted in red. \textbf{c,} The Lorentzian fitting of the momentum distribution curve at $E_{\textrm{B}}=0.25$ eV. \textbf{d,} The dispersion data points ($E_{\textrm{B}}$, $k$) of the surface states are obtained from the Lorentzian fitting of the momentum distribution curves at different binding enerigies and shown by the solid blue circles. Then the surface state dispersion is fitted by linear function in order to obtain the velocity of the surface states.}
\end{figure*}

\clearpage
\begin{figure*}[h]
\includegraphics[width=16cm]{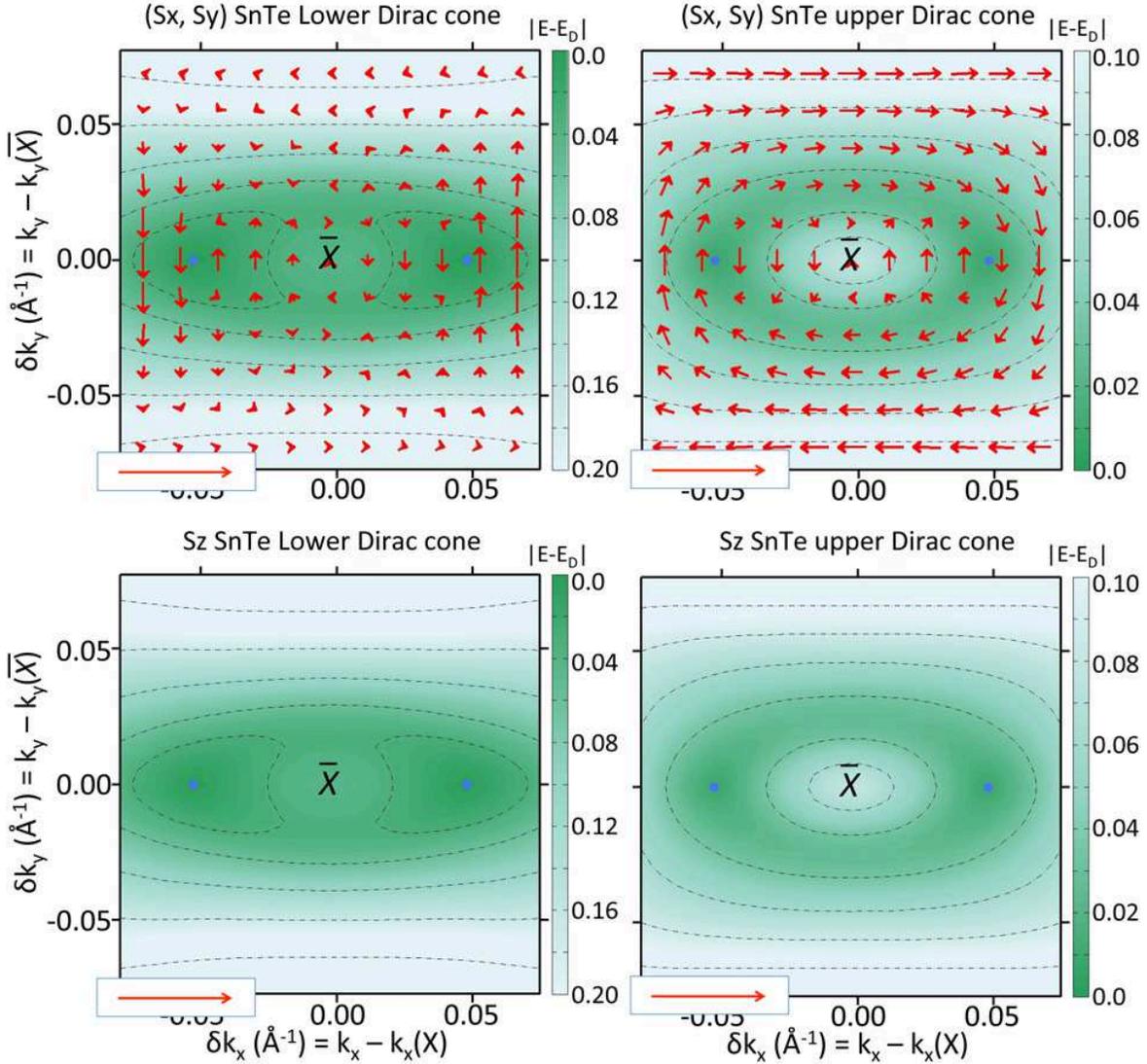}
\caption{\label{Spin_cal_tot} \textbf{Spin texture of the (001) SnTe surface states in the vicinity of an $\bar{X}$ point based on first-principles calculation.} The in-plane components $(P_x, P_y)$ of the SnTe surface states spin texture. The red arrows represent the spin polarization at the corresponding momentum space location. The direction and the length of the arrows show the direction and the magnitude of the spin polarization, respectively. The insets represent the length of the unit spin vector with spin polarization of 1 (100\%). The out-of-plane component is found to be zero throughout the calculated $(E_{\textrm{B}}, k_x, k_y)$ range. The two blue dots denote the two Dirac points on the opposite sides of the $\bar{X}$ point. The color scale shows the energy of the surface states' contours with respect to the Dirac point energy ($E_{\textrm{D}}$). The black dotted lines show the iso-energetic contours of the surface states at different energies. The left panel shows the spin texture below the Dirac points (lower Dirac cones), and the right panel shows the spin texture above the Dirac points (upper Dirac cones).}
\end{figure*}

\clearpage
\begin{figure*}[h]
\includegraphics[width=18cm]{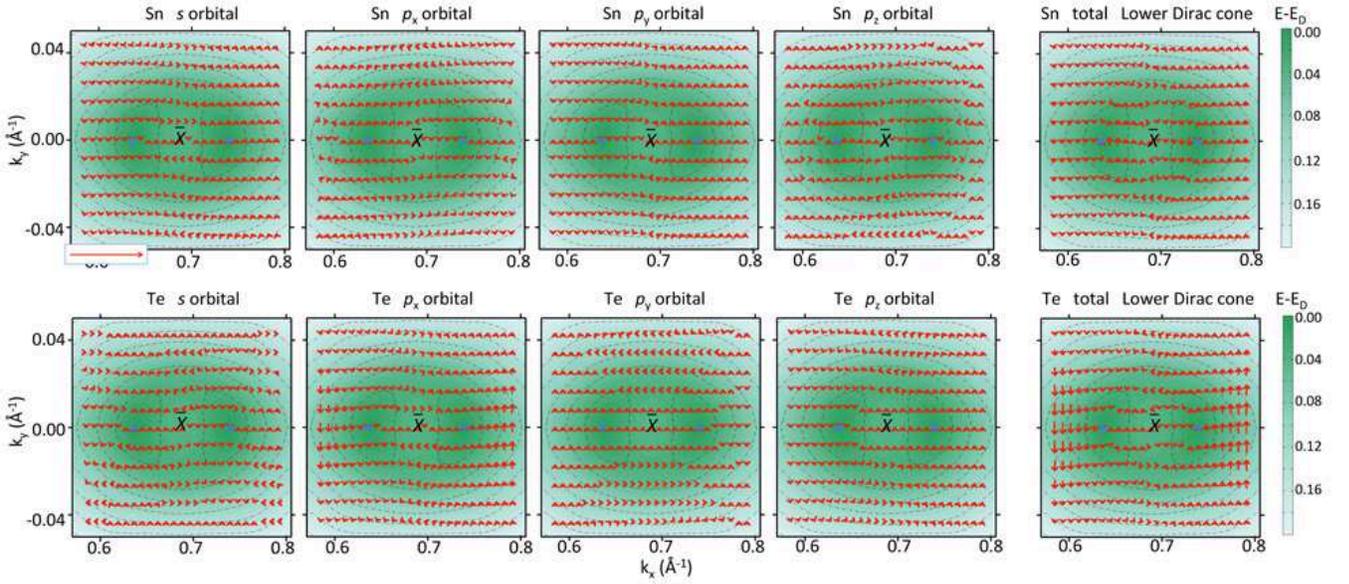}
\caption{\label{Spin_cal_orb} \textbf{Calculated spin texture decomposed into their orbital origins.} The spin of $s$, $p_x$, $p_y$, and $p_z$ orbitals of Sn and Te atoms are shown for the lower Dirac cones of the surface states. The inset represents the length of the unit spin vector with spin polarization of 1 (100\%) for all panels. The two blue dots denote the two Dirac points on the opposite sides of the $\bar{X}$ point. The color scale shows the energy of the surface states' contours with respect to the Dirac point energy ($E_{\textrm{D}}$). The black dotted lines show the iso-energetic contours of the surface states at different energies.}
\end{figure*}

\clearpage
\begin{figure*}[h]
\includegraphics[width=17cm]{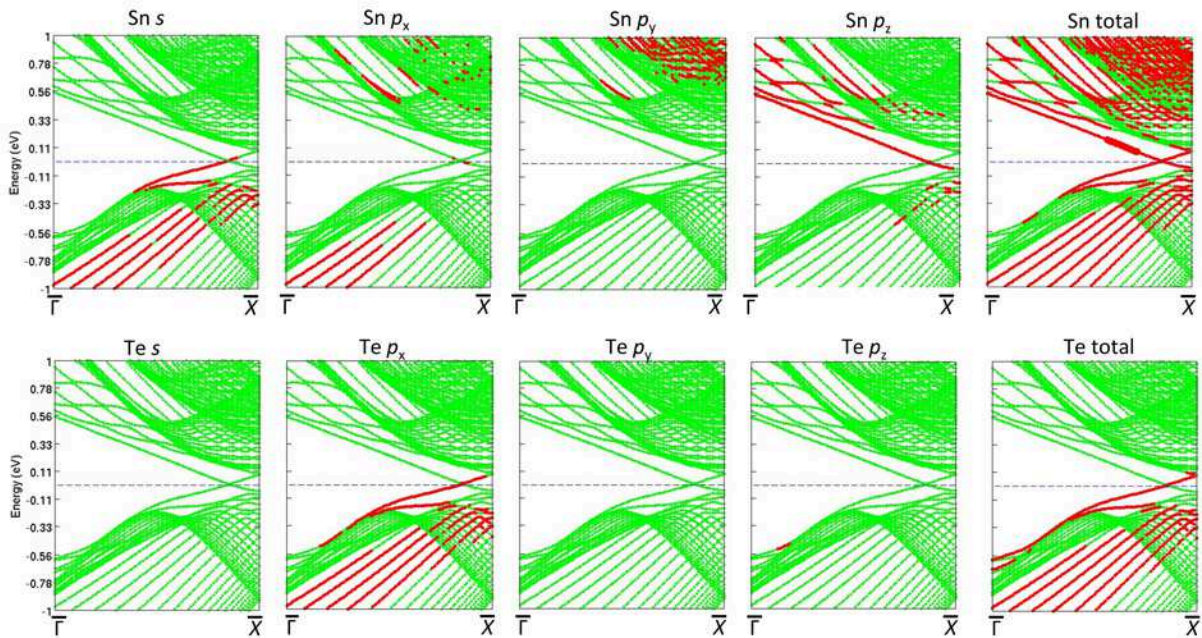}
\caption{\label{Orb} \textbf{Orbital projections of the SnTe (001) surface states.} The red lines denote that the corresponding electronic states have a strong contribution (weight) from certain orbitals, whereas the green lines denote that the corresponding electronic states have a very small contribution (weight) from certain orbitals.}
\end{figure*}

\begin{figure*}[h]
\includegraphics[width=15cm]{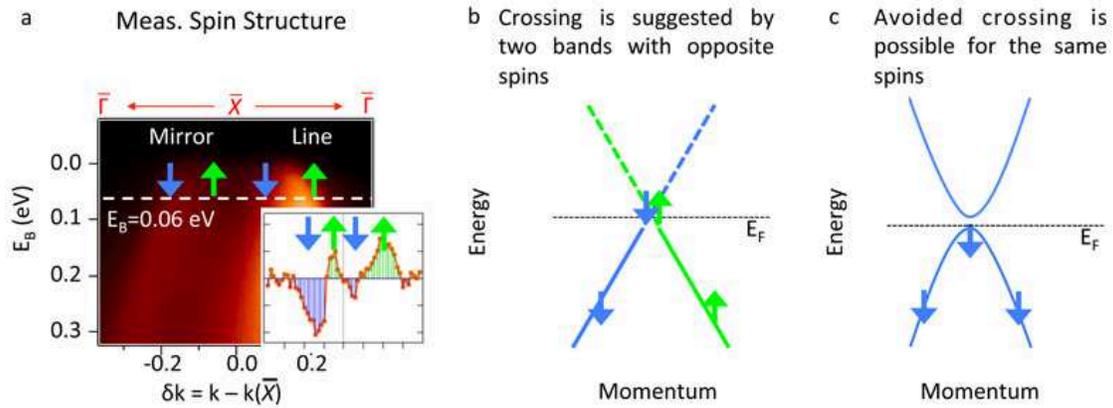}
\centering
\caption{\label{Crossing} \textbf{Spin polarization profile and Dirac crossing.} \textbf{a,} The spin polarization of the surface states along the $\bar{\Gamma}-\bar{X}-\bar{\Gamma}$ mirror line direction revealed by our spin-resolved measurement SR-Cut 1. \textbf{b,} Theoretically, two spin singlet bands with opposite spin directions can only cross each other. \textbf{c,} Avoided crossing is allowed for band with the same spin flavor, or a spin degenerate band.}
\end{figure*}

\clearpage
\begin{figure*}[h]
\begin{center}
\includegraphics[width=17cm]{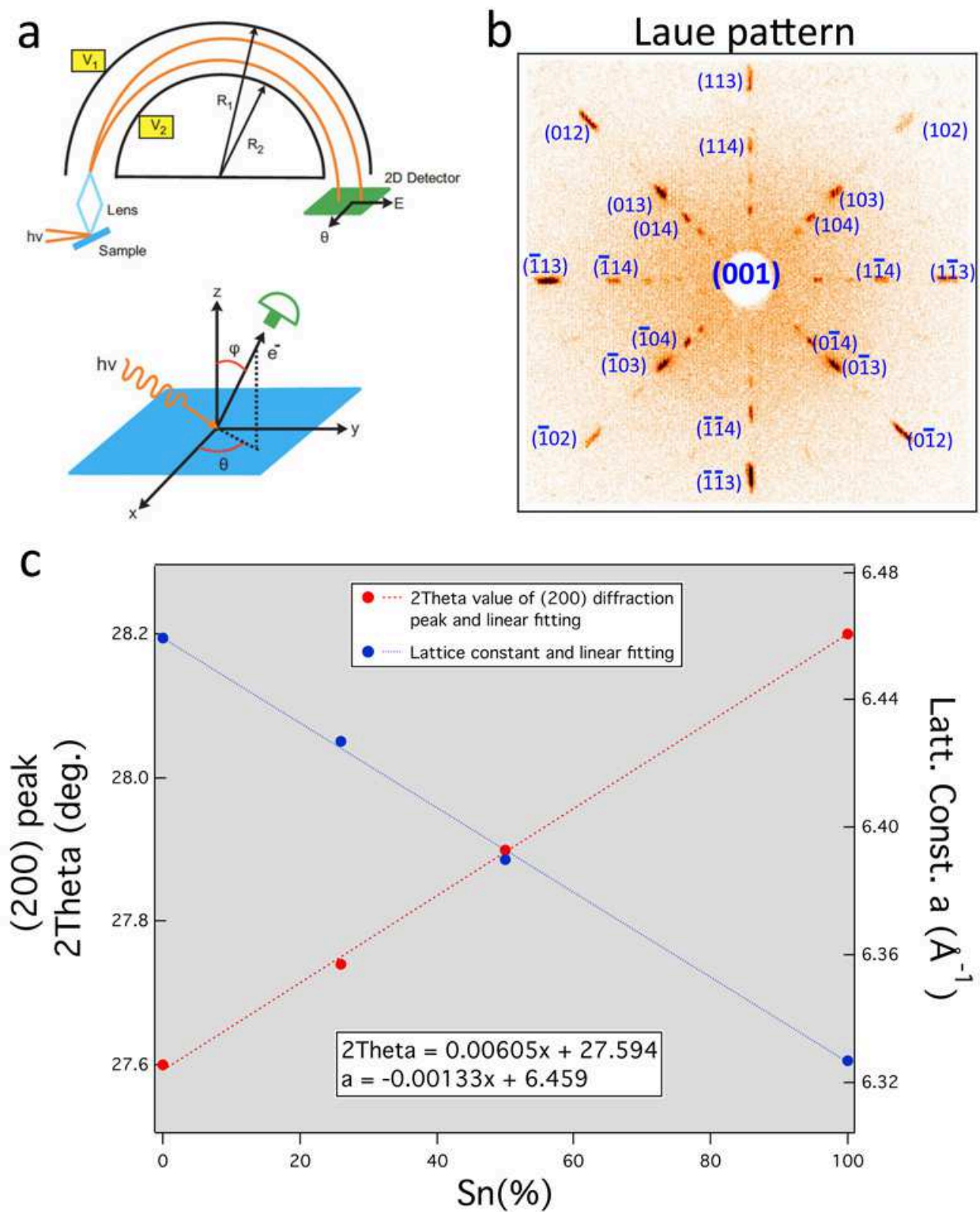}
\caption{\label{ARPES} \textbf{Scattering geometry and sample characterization.} \textbf{a,} Top: The electron analyzer setup used to measure emitted photoelectrons from the sample surface.}
\end{center}
\end{figure*}
\addtocounter{figure}{-1}
\begin{figure} [t!]
\caption{(Previous page.) The two hemispheres detect the electrons to produce a 2D image of energy vs momentum in one shot. Bottom: The geometry of the detector relative to the sample surface. The momentum of the electron inside the sample can be extracted from the measured values of $\textrm{E}_{\textrm{kin}}$, $\theta$ and $\phi$. \textbf{b,} X-ray Laue diffraction pattern of a representative Pb$_{0.6}$Sn$_{0.4}$Te sample used for ARPES experiments. The Miller indices are noted for the diffraction peaks in Laue pattern, which reveals that the cleavage surface of the crystal is perpendicular to the [001] vertical crystal axis. \textbf{c,} The 2Theta values of the (200) Bragg peak (the sharpest and most intensive peak for X-ray diffraction on Pb$_{1-x}$Sn$_{x}$Te \cite{PbTe, Sn26}, also see Supplementary Figure ~\ref{XRD}a) and the corresponding lattice constant values for PbTe ($x=0\%$), Pb$_{0.74}$Sn$_{0.26}$Te ($x=26\%$), Pb$_{0.50}$Sn$_{0.50}$Te ($x=50\%$), and SnTe ($x=100\%$) obtained from  \cite{PbTe, Sn26, Sn50, SnTe} respectively. Linear fittings are applied to the data points, from which we obtain $2\textrm{Theta}=0.00605x+27.594$ and $a=-0.00133x+6.459$.}
\end{figure}

\clearpage
\begin{figure*}[h]
\centering
\includegraphics[width=13.5cm]{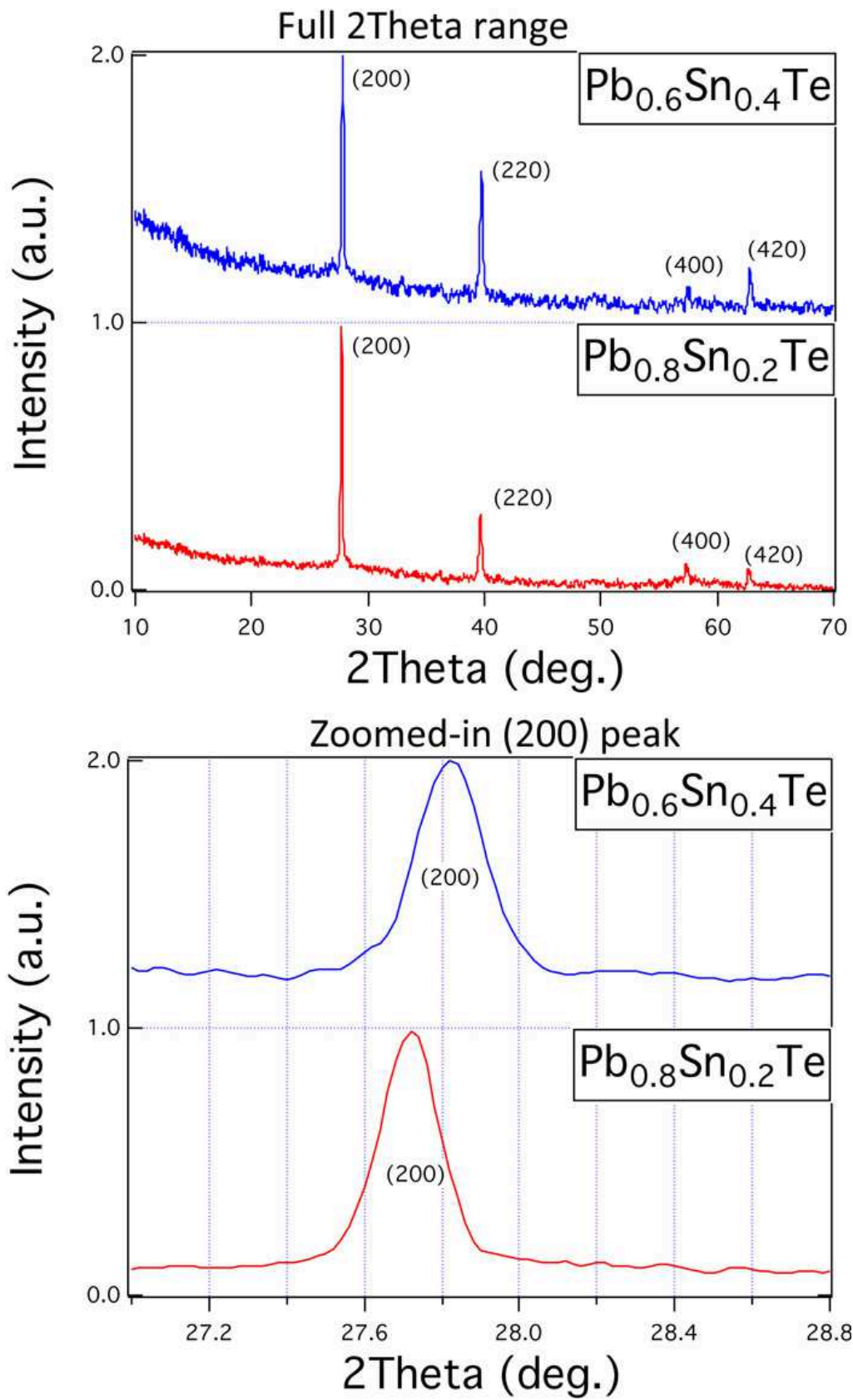}
\caption{\label{XRD} \textbf{X-ray diffraction measurements on two representative compositions.}}
\end{figure*}
\addtocounter{figure}{-1}
\begin{figure} [t!]
\caption{(Previous page.)  \textbf{a,} X-ray diffraction measurements on the two representative compositions. The intensity of the XRD measurements is normalized (maximum intensity to be 1). The XRD data of Pb$_{0.6}$Sn$_{0.4}$Te is offset by 1 with respect to that of Pb$_{0.8}$Sn$_{0.2}$Te.  The corresponding indices of the Bragg peaks are noted. \textbf{b,} The diffraction data zoomed-in in close vicinity of the (200) Bragg peak for the two compositions. The 2Theta values of the (200) Bragg peak are found to be 27.73 deg and 27.83 deg for the two compositions. Using the formula $2\textrm{Theta}=0.00605x+27.594$ obtained from linear fittings in Supplementary Figure~\ref{ARPES}c, we obtain that these 2Theta values correspond to x values of $x=0.22$ (for the nominal concentration $x_{\textrm{nom}}=0.20$ samples) and $x=0.39$ (for the nominal concentration $x_{\textrm{nom}}=0.36$ samples). In the paper we note the chemical compositions as Pb$_{0.8}$Sn$_{0.2}$Te and Pb$_{0.6}$Sn$_{0.4}$Te.}
\end{figure}

\clearpage

\textbf{\large {Supplementary Discussion}}

\begin{itemize}
\bigskip
\item {\textbf{The topological distinction between TI  and TCI phases}}

\begin{itemize}
\item \textbf{\underline{TCI vs. TI / Even vs. Odd number of band inversions}:} It is important to note that the Pb$_{1-x}$Sn$_x$Te system contains an even number of inversions [9]. Traditionally even number of band inversions are believed to be topologically trivial, and only odd number of inversions are believed to give rise to topological insulator states [2]. For this reason, although the Pb$_{1-x}$Sn$_x$Te system has long been known to contain an even number of inversions, they were believed to be trivial insulators. (see Ref. [2]). Since the much studied Z$_2$ TIs can only be realized via odd number of inversions, thus the even number inversions in the SnTe system determines that such system is topologically distinct from both the TI phase and the trivial band insulator phase, which means TCI is a new phase of topological matter. 

\item \textbf{\underline{Irrelevance of Time-Reversal Symmetry Protection}:} The topological distinction between TCI and TI can be directly seen from our ARPES measured surface states. For example, Fig. 5 in the maintext shows an ARPES comparison between a Z$_2$ (Kane-Mele) TI GeBi$_2$Te$_4$ and our new Pb$_{0.6}$Sn$_{0.4}$Te samples. As shown in Fig. 5 of the maintext, GeBi$_2$Te$_4$ belongs to the well-understood single Dirac cone topological insulator, which is topologically the same as the famous Bi$_2$Se$_3$ or Bi$_2$Te$_3$ systems. GeBi$_2$Te$_4$ has only one surface state (odd number), which is found to enclose the time-reversal invariant momenta (TRIM) or the KramersÕ point (the $\bar{\Gamma}$  point). For this reason, the single Dirac cone surface state in GeBi$_2$Te$_4$ (also Bi$_2$Se$_3$, Bi$_2$Te$_3$, and all other Z$_2$ TI systems) is topologically protected by time-reversal symmetry. In sharp contrast, in the Pb$_{0.6}$Sn$_{0.4}$Te system, none of the surface states are found to enclose any of the time-reversal invariant momenta (TRIM). Thus the surface states in the Pb$_{0.6}$Sn$_{0.4}$Te system are irrelevant to time-reversal symmetry type of protection. The topological protection in the case of Pb$_{0.6}$Sn$_{0.4}$Te is a result of the mirror (spatial) symmetries of the crystal, as theoretically predicted by Ref. [15]. One relevant experimental evidence for the mirror protection is that all of the observed surface states are located on the mirror line momentum space directions ($\bar{\Gamma}-\bar{X}-\bar{\Gamma}$).
\end{itemize}

\item \textbf{Spin polarization, Dirac crossing and gapless nature}

Here we provide one experimental observation relevant to the Dirac crossing and gapless nature: the spin polarization profile. As shown in Supplementary Figure~\ref{Crossing}, for two electronic bands with opposite spin directions, crossing is the only fate. The avoided crossing and gap behavior is only allowed if the bands have the same spin flavor or they are spin degenerate. Here we show that at energy level 60 meV below the chemical potential (thus about $60\pm20$ meV around the Dirac point), the two branches of each surface state cone show the expected opposite spin directions. It has been recently experimentally shown that the spin texture of the Dirac surface states will be strongly disturbed when a gap is opened at the Dirac point [38]. The energy scale where the spin texture is disturbed roughly corresponds to the gap value [38]. Therefore, by showing a helical spin structure at about $60\pm20$ meV around the Dirac point in Supplementary Figure~\ref{Spin_cut1}c and Supplementary Figure~\ref{Crossing}a, we can safely exclude a gap value of $50$ meV. However, whether there is a ${\leq}$ 10 meV gap or not in our surface states still need further studies.
\end{itemize}

\newpage
\textbf{\large {Supplementary Methods}}
\begin{itemize}
\item \textbf{Laue measurements}

The surface cleavage termination is determined by X-ray diffraction measurements on the cleaved samples after ARPES measurements. Supplementary Figure~\ref{ARPES}b shows a representative Laue measurement on the Pb$_{0.6}$Sn$_{0.4}$Te ($x=0.4$) sample. The Miller indices are noted for the diffraction peaks in Laue pattern, which reveals that the cleavage surface of the crystal is perpendicular to the [001] vertical crystal axis. The (001) surface termination determined here from Laue results is consistent with the four-fold symmetry of the ARPES Fermi surfaces and the size of the surface BZ in ARPES measurements.

\item \textbf{X-ray diffraction measurements}

As shown in Supplementary Figure~\ref{XRD} for our XRD measurements on the two representative compositions, the 2Theta values of the (200) Bragg peak are found to be 27.73 deg (for the nominal concentration $x_{\textrm{nom}}=0.20$ samples, noted as Pb$_{0.8}$Sn$_{0.2}$Te in the maintext) and 27.83 deg (for the nominal concentration $x_{\textrm{nom}}=0.36$ samples, noted as Pb$_{0.6}$Sn$_{0.4}$Te in the maintext). Using the linear fitting above, these 2Theta values correspond to x values of $x=0.22$ (for the nominal concentration $x_{\textrm{nom}}=0.20$ samples) and $x=0.39$ (for the nominal concentration $x_{\textrm{nom}}=0.36$ samples). In the paper we note the chemical compositions as Pb$_{0.8}$Sn$_{0.2}$Te and Pb$_{0.6}$Sn$_{0.4}$Te.

\item \textbf{Incident photon energy dependence study on Pb$_{1-x}$Sn$_{x}$Te}


As a qualitative guide to the ARPES incident photon energy dependence measurements, here we show the first-principles calculated bulk bands $k_z$ dispersion of the two end compounds inverted SnTe (Supplementary Figure~\ref{SnTe_BVB}) and non-inverted PbTe (Supplementary Figure~\ref{PbTe_BVB}) respectively.

Now we present comparative incident photon energy dependence studies between the inverted Pb$_{0.6}$Sn$_{0.4}$Te samples and the non-inverted Pb$_{0.8}$Sn$_{0.2}$Te samples under the identical experimental conditions and setups (beamline, incident photon energy values, incident light polarization, sample surface preparation procedure, etc.) as shown in Supplementary Figure~\ref{hv_dep_com}. A pair of surface states without observable $k_z$ dispersion is observed in the inverted $x=0.4$ samples but absent in the non-inverted $x=0.2$ samples. no surface state on the Fermi level is observed for all the incident photon energy applied under the same experimental conditions and setups as shown in Supplementary Figure~\ref{hv_dep_com}a. The $k_z$ evolution (dispersion) of the bulk valence band at different $k_z$ values (different incident photon energies) is in qualitative agreement with the theoretical calculation results shown above. The incident photon energy range from 26 eV to 10 eV shown in Supplementary Figure~\ref{hv_dep_com} corresponds to a wide $k_z$ range from $1.8\pi$ to $0.1\pi$, which corresponds to going from the top of the bulk BZ to the center of the bulk BZ.

\item \textbf{Fitting results of the experimental chemical potential with respect to the Dirac point energy}

We apply a linear fitting to the ARPES dispersion near the Fermi level along Cut 2 in the maintext, which enables us to extract the velocity of the surface states, as well as the chemical potential ($E_{\textrm{F}}$) with respect to the Dirac point energy ($E_{\textrm{D}}$). 

\begin{itemize}
\item Supplementary Figure~\ref{Fitting}a shows the Fermi surface mapping zoomed-in in the vicinity of an $\bar{X}$ point. The blue dotted line defines a momentum space cut direction (which is the same as Cut 2 in Fig. 3 of the maintext).
\item Supplementary Figure~\ref{Fitting}b shows the dispersion map and the corresponding momentum distribution curves (MDCs) along the momentum space cut defined by the blue dotted line in Supplementary Figure~\ref{Fitting}a.
\item The MDC is the ARPES measured intensity distribution along the momentum axis at a fixed binding energy $E_{\textrm{B}}$. For example, the white solid line in Supplementary Figure~\ref{Fitting}b is at binding energy $E_{\textrm{B}}=0.25$ eV and the corresponding MDC is highlighted in red in the MDC panel in Supplementary Figure~\ref{Fitting}b.
\item We fit the MDCs by two Lorentzian peaks. We take the MDC Lorentzian fitting at $E_{\textrm{B}}=0.25$ eV as an example, as shown in Supplementary Figure~\ref{Fitting}c. The Lorentzian peak position reveals the momentum location of each branch of the Dirac surface states at $E_{\textrm{B}}=0.25$ eV.
\item By performing such fitting at different binding energies, we obtain the dispersion of the surface states, namely ($E_{\textrm{B}}$, $k$). The solid blue circles in Supplementary Figure~\ref{Fitting}E show the obtained dispersion data points of the surface states from $E_{\textrm{B}}=0.25$ eV to $E_{\textrm{B}}=0.05$ eV with a binding energy step of ${\Delta}E_{\textrm{B}}=0.02$ eV. 
\item Then we fit the surface states' dispersion data points by a linear function. For the two branches of the surface state cone we obtain the following: for the branch with positive slope, we have $E_{\textrm{B}}=(2.68(\pm0.07)\textrm{eV}{\cdot}\textrm{\AA}){\cdot}k+0.008(\pm0.009)\textrm{eV}$. And for branch with negative slope, we have $E_{\textrm{B}}=(-2.88(\pm0.08)\textrm{eV}{\cdot}\textrm{\AA}){\cdot}k-0.010(\pm0.008)\textrm{eV}$. Thus, the linear fittings of both branches give the velocity (the absolute value of the slope) of the Dirac cone to be around $2.8$ $\textrm{eV}{\cdot}{\textrm{\AA}}$. The Dirac point energy $E_{\textrm{D}}$ is given by the binding energy intercept ($E_{\textrm{B}}(k=0)$) of the linear fitting, which is 0.008 eV and -0.010 eV for each fitting respectively. This means the Dirac point energy $E_{\textrm{D}}$ implied by the fitting of the positive slope branch is 0.008 eV below the chemical potential $E_{\textrm{F}}$, whereas the Dirac point energy $E_{\textrm{D}}$ implied by the fitting of the negative slope branch is 0.010 eV above the chemical potential $E_{\textrm{F}}$.
\item In the maintext, we report a surface state velocity of $2.8$ $\textrm{eV}{\cdot}{\textrm{\AA}}$ along this momentum space cut direction (Cut 2 in Fig. 3 of the maintext). Furthermore the chemical potential with respect to the Dirac point is found to be $E_{\textrm{F}}=E_{\textrm{D}}\pm0.02$ eV
\end{itemize}

\item \textbf{Spin-resolved measurements on Pb$_{0.6}$Sn$_{0.4}$Te}

Here we show systematic spin dataset of these measurements for all three components of the spin polarization vector. The $\hat{x}$, $\hat{y}$ directions for spin polarization vectors are defined in the inset of Supplementary Figure~\ref{Spin_cut1}a. And the $\hat{z}$ direction follows the right-hand rule. The $P_x$, $P_y$ and $P_z$ measurements for SR-Cut 1 and SR-Cut 2 are shown in Supplementary Figure~\ref{Spin_cut1} and Supplementary Figure~\ref{Spin_cut2}. As discussed in the main paper, in total four in-plane tangential ($P_y$) spins are revealed by these polarization measurements (Supplementary Figure~\ref{Spin_cut1}c) on the surface states. We note that the magnitude of the spin polarization is between $10\%-20\%$, which is smaller than the typically observed 40\% net spin polarization of the Z$_2$ topological insulator surface states \cite{David Nature tunable}. In fact the $10\%-20\%$ spin polarization in our $x=0.4$ samples is found to be consistent with the first-principles based spin texture calculation on SnTe surface states (see below). As shown in Supplementary Figure~\ref{Spin_cut1}c, the polarization magnitude of the inner two branches are found to be smaller as compared to that of the outer two branches. This can be understood by two independent reasons: First, since the inner two branches are close to each other in momentum space, their polarization signal can intermix with each other due to the finite momentum resolution of the spin-resolved measurements (the momentum resolution is about 3\% of the surface BZ which corresponds to roughly $0.04\textrm{\AA}^{-1}$). Second, the smaller polarization of the inner two branches is even observed in our calculation on SnTe (see below). This effect is related to the anisotropy of the surface states and the hybridization of the two adjacent Dirac cones.

\item \textbf{Spin texture calculation of SnTe}

The spin texture of SnTe is calculated using VASP first-principles method. We have assumed the SnTe lattice in ideal sodium chloride structure without rhombohedral distortion. The calculated spin texture of the SnTe surface states is shown in Supplementary Figure~\ref{Spin_cal_tot} for the lower and upper Dirac cone, respectively. The out-of-plane component of the spin texture is found to be zero. We compare the calculated spin texture with the spin polarization measurements shown above: Along the $\bar{\Gamma}-\bar{X}-\bar{\Gamma}$ mirror line direction ($k_y=0$ in Supplementary Figure~\ref{Spin_cal_tot}a), both theoretical calculations in Supplementary Figure~\ref{Spin_cal_tot}a (low Dirac cone) and spin measurements in Supplementary Figure~\ref{Spin_cut1}c reveal in total 4 spins with alternating directions in going along the mirror line. The magnitude of the spin polarization in calculation is found to range $10\%$ to $30\%$, which is consistent with the experiments in Supplementary Figure~\ref{Spin_cut1}c. The inner two spins close to the $\bar{X}$ point are found to have smaller polarization as compared to the outer two, which is also in agreement with the experimental results in Supplementary Figure~\ref{Spin_cut1}c.

In order to better understand the spin texture, we also calculate the atomic orbital contribution of the spin texture. For the (001) surface, as shown in Supplementary Figure~\ref{Orb}. 
\end{itemize}


\begin{thebibliography}{21} 
\bibitem{Moore} Hasan, M. Z. \& Moore, J. E. Three-Dimensional Topological Insulators. \textit{Ann. Rev. Cond. Mat. Phys.} $\mathbf{2}$, 55-78 (2010).

\bibitem{RMP} Hasan, M. Z. \& Kane, C. L. Topological insulators. \textit{Rev. Mod. Phys.} $\mathbf{82}$, 3045-3067 (2010).
\bibitem{Zhang_RMP} Qi, X. -L. \& Zhang, S. -C. Topological insulators and superconductors. \textit{Rev. Mod. Phys.} $\mathbf{83}$, 1057-1110 (2011).
\bibitem{Matthew Nature physics BiSe} Xia, Y. \textit{et al}. Observation of a large-gap topological-insulator class with a single Dirac cone on the surface. \textit{Nature Phys.} $\mathbf{5}$, 398-402 (2009).
\bibitem{Hasan QPT} Xu, S. -Y. \textit{et al}. Topological phase transition and texture inversion in a tunable topological insulator. \textit{Science} $\mathbf{332}$, 560-564 (2011).


\bibitem{PbTe IR} Zogg, H., Fach, A., Masek, J. \& Blunier, S. Photovoltaic lead-chalcogenide on silicon infrared sensor arrays. \textit{Opt. Eng.} $\mathbf{33}$, 1440-1449 (1994). 
\bibitem{PbTe Thermal} Zhu, P. W. \textit{et al}. Thermoelectric properties of PbTe prepared at high pressure and high temperature. \textit{J. Phys.: Condens. Matter} $\mathbf{14}$, 11185-11188 (2002).

\bibitem{PST band-gap} Dimmock, J. O. \& Wright, G. B. Band Edge Structure of PbS, PbSe, and PbTe. \textit{Phys. Rev.} $\mathbf{135}$, 821-830 (1964).
\bibitem{PST Inversion1} Dimmock, J.O., Melngailis, I. \& Strauss, A.J. Band structure and laser action in Pb$_x$Sn$_{1-x}$Te. \textit{Phys. Rev. Lett.} $\mathbf{16}$, 1193-1196 (1966).
\bibitem{PST Inversion2} Gao, X. \& Daw, M. S. Investigation of band inversion in (Pb,Sn)Te alloys using ab initio calculations. \textit{Phy. Rev. B} $\mathbf{77}$, 033103 (2008).
\bibitem{PST Inversion3} Pankratov, O. A., Pakhomov, S. V. \& Volkov, B. A. Supersymmetry in heterojunctions: Band-inverting contact on the basis of Pb$_{1-x}$Sn$_{x}$Te and Hg$_{1-x}$Cd$_x$Te. \textit{Solid State Commun.} $\mathbf{61}$, 93-96 (1987).
\bibitem{Volkov} Volkov, B. A. \& Pankratov, O. A. Two-dimensional massless electrons in an inverted contact. \textit{JETP Lett.} $\mathbf{42}$, 178-181 (1985).
\bibitem{Fradkin} Fradkin, E., Dagotto, E. \& Boyanovsky, D. Physical Realization of the Parity Anomaly in Condensed Matter Physics. \textit{Phys. Rev. Lett.} $\mathbf{57}$, 2967-2970 (1986). 

\bibitem{Liang PRL TCI} Fu, L. Topological Crystalline Insulators. \textit{Phys. Rev. Lett.} $\mathbf{106}$, 106802 (2011).
\bibitem{Liang NC SnTe} Hsieh, H. \textit{et al}. Topological Crystalline Insulators in the SnTe Material Class. \textit{Nature Comm.} $\mathbf{3}$, 982 (2012).

\bibitem{PST Rhombohedral Distortion} Iizumi, M. \textit{et al}. Phase Transition in SnTe with Low Carrier Concentration. \textit{J. Phys. Soc. Jpn.} $\mathbf{38}$, 443-449 (1975).

\bibitem{SnTe p-type} Burke, Jr., J. R., Allgaier, R. S., Houston, Jr., B. B., Babiskin, J. \& Siebenmann, P. G. Shubnikov-de Haas effect in SnTe. \textit{Phys. Rev. Lett.} $\mathbf{14}$, 360-361 (1965).
\bibitem{SnTe ARPES} Littlewood, P. B. \textit{et al}. Band structure of SnTe studied by Photoemission Spectroscopy. \textit{Phys. Rev. Lett.} $\mathbf{105}$, 086404 (2010). 

\bibitem{PST n-type} Takafuji Y. \& Narita S. Shubnikov-de Haas Measurements in N-Type Pb$_{1-x}$Sn$_x$Te. \textit{Jpn. J. Appl. Phys.} $\mathbf{21}$, 1315-1322 (1982).
\bibitem{Yannopapas} Yannopapas, V. Gapless surface states in a lattice of coupled cavities: A photonic analog of topological crystalline insulators. \textit{Phys. Rev. B} $\mathbf{84}$, 195126 (2011).
\bibitem{Wang} Hao, N., Zhang, P. \& Wang, Y. Topological phases and fractional excitations of the exciton condensate in a special class of bilayer systems. \textit{Phys. Rev. B} $\mathbf{84}$, 155447 (2011).
\bibitem{Vildanov} Vildanov, N. M. Effective field theory description of topological crystalline insulators. Preprint at http://arXiv.org/abs/1205.3560 (2012).

\bibitem{David Nature BiSb} Hsieh, D. \textit{et al}. A topological Dirac insulator in a quantum spin Hall phase. \textit{Nature} $\mathbf{452}$, 970-974 (2008).

\bibitem{Liang PRL Warping} Fu, L. Hexagonal warping effects in the surface states of the topological insulator Bi$_2$Te$_3$.  \textit{Phys. Rev. Lett.} $\mathbf{103}$, 266801 (2009).




\bibitem{Spin1} Hoesch, M. \textit{et al}. Spin-polarized Fermi surface mapping. \textit{J. Electron Spectrosc. Relat. Phenom.} $\mathbf{124}$, 263-279 (2002).
\bibitem{Spin2} Dil, J. H. \textit{et al}. Spin and angle resolved photoemission on non-magnetic low-dimensional systems. \textit{J. Phys. Condens. Matter} $\mathbf{21}$, 403001 (2009).

\bibitem{Mirror Chern Number} Teo, J. C. Y., Fu, L. \& Kane, C. L. Surface states and topological invariants in three-dimensional topological insulators: Application to Bi$_{1-x}$Sb$_x$. \textit{Phys. Rev. B} $\mathbf{78}$, 045426 (2008).
\bibitem{David Science BiSb} Hsieh, D. \textit{et al}. Observation of Unconventional Quantum Spin Textures in Topological Insulators. \textit{Science} $\mathbf{323}$, 919-922 (2009).

\bibitem{PSS TCI} Dziawa, P. \textit{et al}. Topological crystalline insulator states in Pb$_{1-x}$Sn$_x$Se. Preprint at arXiv:1206.1705v1
\bibitem{PST Xu} Xu, S. -Y. \textit{et al}. Observation of Topological Crystalline Insulator phase in the lead tin chalcogenide Pb$_{1-x}$Sn$_x$Te material class. Preprint at http://arXiv.org/abs/1206.2088 (2012).

\bibitem{Ternary arXiv} Xu, S. -Y. \textit{et al}. Discovery of several large families of Topological Insulator classes with backscattering-suppressed spin-polarized single-Dirac-cone on the surface. Preprint at http://arXiv.org/abs/1007.5111 (2010).
\bibitem{Ternary PRB} Neupane, M. \textit{et al}. Topological surface states and Dirac point tuning in ternary topological insulators. \textit{Phys. Rev. B} $\mathbf{85}$, 235406 (2012).
\bibitem{Kimura} Okamoto, K. \textit{et al}. Observation of a Highly Spin Polarized Topological Surface State in GeBi$_{2}$Te$_{4}$. Preprint at http://arXiv.org/abs/1207.2088 (2012).



\bibitem{PbMnTe} Story, T. \textit{et al}. Carrier-concentration-induced ferromagnetism in PbSnMnTe. \textit{Phys. Rev. Lett.} $\mathbf{56}$, 777-779 (1986) .
\bibitem{Tl-PbTe ARPES} Nakayama, K., Sato, T., Takahashi, T. \& Murakami, H. Doping Induced Evolution of Fermi Surface in Low Carrier Superconductor Tl-Doped PbTe. \textit{Phys. Rev. Lett.} $\mathbf{100}$, 227004 (2008).
\bibitem{In-SnTe Ando} Sasaki, S. \textit{et al}. Odd-Parity Pairing and Topological Superconductivity in a Strongly Spin-Orbit
Coupled Semiconductor. Preprint at http://arXiv.org/abs/1208.0059 (2012).

\bibitem{Andrew CuBiSe} Wray, L. A. \textit{et al}. Observation of topological order in a superconducting doped topological insulator. \textit{Nature Phys.} $\mathbf{6}$, 855-859 (2010).
\bibitem{Hedgehog} Xu, S.-Y. \textit{et al}. Hedgehog spin texture and Berry's phase tuning in a magnetic topological insulator.  \textit{Nature Phys.} $\mathbf{8}$, 616-622 (2012).



\bibitem{Growth} Nugraha, K. Suto, O. Itoh, J. Nishizawa, Y. Yokota, Growth and electrical properties of PbTe bulk crystals grown by the Bridgman method under controlled tellurium or lead vapor pressure. \textit{J. Cryst. Gr.} $\mathbf{165}$, 402-407 (1996).

\bibitem{PAW} Kresse, G. \& Joubert, D. From ultrasoft pseudopotentials to the projector augmented-wave method. \textit{Phys. Rev. B} $\mathbf{59}$, 1758-1775 (1999).
\bibitem{PBE} Perdew, J. P., Burke, K. \& Ernzerhof, M. Generalized Gradient Approximation Made Simple. \textit{Phys. Rev. Lett.} $\mathbf{77}$, 3865-3868 (1996).
\bibitem{VASP} Kresse, G. \& Hafner, J., Ab initio molecular dynamics for open-shell transition metals. \textit{Phys. Rev. B} $\mathbf{48}$, 13115-13118 (1993).
\bibitem{LSalts} Bis, R. F \& Dixon, J. R. Applicability of Vagard's Law to the Pb$_x$Sn$_{1-x}$Te Alloy System. \textit{J. Appl. Phys.} $\mathbf{40}$, 1918-1921 (1969).


\bibitem{PbTe} Bouad, N. \textit{et al}. Neutron powder diffraction study of strain and crystallite size in mechanically alloyed PbTe. \textit{J. Solid State Chem.} $\mathbf{173}$, 189-195 (2003).
\bibitem{Sn26} Ishida, A., Aoki, M. Fujiyasu, H. Sn diffusion effects on x-ray diffraction patterns of Pb$_{1-x}$Sn$_x$Te-PbSe$_y$Te$_{1-y}$superlattices. \textit{J. Appl. Phys.} $\mathbf{58}$, 797-801 (1985).
\bibitem{Sn50} Yakimova, R. T., Trifonova, E. P., Karagiozov, L. \& Petrov, S. Structural and electrical characteristics of iodidely synthesized Pb$_{1-x}$Sn$_x$Te crystals. \textit{S. Cryst. Res. Technol.} $\mathbf{19}$, K109-K112 (1984).
\bibitem{SnTe} Scheer, M., McCarthy, G., Seidler, D. \& Boudjouk, P. North Dakota State Univ., Fargo, ND, USA., ICDD Grant-in-Aid (1994).
\bibitem{David Nature tunable} Hsieh, D. \textit{et al}. A tunable topological insulator in spin helical Dirac transport regime. \textit{Nature} $\mathbf{460}$, 1101-1105 (2009).
\end{thebibliography}
\end{document}